\documentclass{emulateapj}
\usepackage{natbib}
\usepackage{amsmath}
\usepackage{graphicx}
\usepackage{apjfonts}

\begin{document}
\slugcomment{{\sc Accepted to ApJ:} March 7, 2014} 

\newcommand{\Mat}{\Omega_{\rm M}}
\newcommand{\Lam}{\Omega_{\Lambda}}
\newcommand{\Cur}{\Omega_{\rm k}}
\newcommand{\Hn}{H$_0$ = 71.6 km s$^{-1}$ Mpc$^{-1}$}

\newcommand{\deltam}{\Delta m_\mu}
\newcommand{\deltamm}{\Delta m_{\mu{\rm, MLCS2k2}}}
\newcommand{\deltams}{\Delta m_{\mu{\rm, SALT2}}}
\newcommand{\dm}{{\rm dm}}
\newcommand{\dmm}{{\rm dm_{MLCS2k2}}}
\newcommand{\dms}{{\rm dm_{SALT2}}}
\newcommand{\dmLCDM}{{\rm dm_{\Lambda CDM}}}
\newcommand{\PIa}{P_{\rm Ia}}
\newcommand{\PIbc}{P_{\rm Ib/c}}
\newcommand{\PII}{P_{\rm II}}

\title{\vskip -0.22in Three Gravitationally Lensed Supernovae Behind CLASH Galaxy Clusters}

\author{ 
Brandon~Patel\altaffilmark{1},
Curtis~McCully\altaffilmark{1},
Saurabh~W.~Jha\altaffilmark{1},
Steven~A.~Rodney\altaffilmark{2,$\dagger$},
David~O.~Jones\altaffilmark{2},
Or~Graur\altaffilmark{2,3,4,5},
Julian~Merten\altaffilmark{6},
Adi~Zitrin\altaffilmark{7,$\dagger$},
Adam~G.~Riess\altaffilmark{2,8},
Thomas~Matheson\altaffilmark{9},
Masao~Sako\altaffilmark{10},
Thomas~W.-S.~Holoien\altaffilmark{1},
Marc~Postman\altaffilmark{8},
Dan~Coe\altaffilmark{8},
Matthias~Bartelmann\altaffilmark{11},
Italo~Balestra\altaffilmark{12,13},
Narciso~Ben\'itez\altaffilmark{14},
Rychard~Bouwens\altaffilmark{15},
Larry~Bradley\altaffilmark{8},
Tom~Broadhurst\altaffilmark{16},
S.~Bradley~Cenko\altaffilmark{17,18},
Megan~Donahue\altaffilmark{19},
Alexei~V.~Filippenko\altaffilmark{18},
Holland~Ford\altaffilmark{2},
Peter Garnavich\altaffilmark{20},
Claudio~Grillo\altaffilmark{21},
Leopoldo~Infante\altaffilmark{22},
St\'ephanie~Jouvel\altaffilmark{23},
Daniel~Kelson\altaffilmark{24},
Anton~Koekemoer\altaffilmark{8},
Ofer~Lahav\altaffilmark{25},
Doron~Lemze\altaffilmark{2},
Dan~Maoz\altaffilmark{3},
Elinor~Medezinski\altaffilmark{2},
Peter~Melchior\altaffilmark{26},
Massimo~Meneghetti\altaffilmark{27},
Alberto~Molino\altaffilmark{14},
John~Moustakas\altaffilmark{28},
Leonidas~A.~Moustakas\altaffilmark{6},
Mario~Nonino\altaffilmark{12},
Piero~Rosati\altaffilmark{29,30},
Stella~Seitz\altaffilmark{31},
Louis~G.~Strolger\altaffilmark{8},
Keiichi~Umetsu\altaffilmark{32},~and
Wei~Zheng\altaffilmark{2}
}

\altaffiltext{1}{Department of Physics and Astronomy, Rutgers, The
  State University of New Jersey, Piscataway, NJ 08854, USA}

\altaffiltext{2}{Department of Physics and Astronomy, The Johns
  Hopkins University, Baltimore, MD 21218, USA}

\altaffiltext{3}{School of Physics and Astronomy, Tel Aviv University,
  Tel Aviv 69978, Israel}

\altaffiltext{4}{Department of Astrophysics, American Museum of
  Natural History, New York, NY 10024, USA}

\altaffiltext{5}{4 CCPP, New York University, 4 Washington Place, New
  York, NY 10003, USA}

\altaffiltext{6}{Jet Propulsion Laboratory, California Institute of
  Technology, MS 169-327, Pasadena, CA 91109, USA}

\altaffiltext{7}{Cahill Center for Astronomy and Astrophysics,
  California Institute of Technology, MS 249-17, Pasadena, CA 91125,
  USA}

\altaffiltext{8}{Space Telescope Science Institute, 3700 San Martin
  Drive, Baltimore, MD 21208, USA}

\altaffiltext{9}{National Optical Astronomy Observatory, 950 North
  Cherry Avenue, Tucson, AZ 85719, USA}

\altaffiltext{10}{Department of Physics and Astronomy, University of
  Pennsylvania, Philadelphia, PA 19104, USA}

\altaffiltext{11}{Institut f\"ur Theoretische Astrophysik,
  Universit\"at Heidelberg, Zentrum f\"ur Astronomie, Philosophenweg
  12, D-69120 Heidelberg, Germany}

\altaffiltext{12}{INAF-Osservatorio Astronomico di Trieste, via
  G.~B.~Tiepolo 11, 34131 Trieste, Italy}

\altaffiltext{13}{INAF-Osservatorio Astronomico di Capodimonte, via
  Moiariello 16, 80131 Napoli, Italy}

\altaffiltext{14}{Instituto de Astrof\'isica de Andaluc\'ia (CSIC),
  Camino Bajo de Hu\'etor 24, Granada 18008, Spain}

\altaffiltext{15}{Leiden Observatory, Leiden University, NL-2300 RA
  Leiden, Netherlands}

\altaffiltext{16}{Department of Theoretical Physics, University of the
  Basque Country, P.~O.~Box 644, 48080 Bilbao, Spain}

\altaffiltext{17}{Astrophysics Science Division, NASA/GSFC, Mail Code
  661, Greenbelt, MD 20771, USA}

\altaffiltext{18}{Department of Astronomy, University of California,
  Berkeley, CA 94720-3411, USA}

\altaffiltext{19}{Department of Physics and Astronomy, Michigan State
  University, East Lansing, MI 48824, USA}

\altaffiltext{20}{Department of Physics, University of Notre Dame, Notre
  Dame, IN 46556, USA}

\altaffiltext{21}{Dark Cosmology Centre, Niels Bohr Institute,
  University of Copenhagen, Juliane Mariesvej 30, DK-2100 Copenhagen,
  Denmark}

\altaffiltext{22}{Institute of Astrophysics, Pontificia Universidad 
Cat\'{o}lica de Chile, Av. Vic. Mackenna 4860, Santiago, Chile}

\altaffiltext{23}{Institut de Ci\'encies de l'Espai (IEEC-CSIC), E-08193
  Bellaterra (Barcelona), Spain}

\altaffiltext{24}{Observatories of the Carnegie Institution of
  Washington, Pasadena, CA 91101, USA}

\altaffiltext{25}{Department of Physics \& Astronomy, University
  College London, Gower Street, London WC1E 6BT, UK}

\altaffiltext{26}{Center for Cosmology and Astro-Particle Physics, \&
  Department of Physics; The Ohio State University, 191 W. Woodruff
  Ave., Columbus, Ohio 43210, USA}

\altaffiltext{27}{INAF-Osservatorio Astronomico di Bologna, \& INFN,
  Sezione di Bologna, Via Ranzani 1, I-40127 Bologna, Italy}

\altaffiltext{28}{Department of Physics and Astronomy, Siena College, 
515 Loudon Road, Loudonville, NY 12211, USA}

\altaffiltext{29}{Dipartimento di Fisica e Scienze della Terra,
  Universit\`a di Ferrara, Via Saragat 1, I-44122 Ferrara, Italy}

\altaffiltext{30}{ESO-European Southern Observatory, D-85748 Garching
  bei M\"unchen, Germany}

\altaffiltext{31}{Universit\"ats-Sternwarte, M\"unchen, Scheinerstr.~1,
  D-81679 M\"unchen, Germany}

\altaffiltext{32}{Institute of Astronomy and Astrophysics, Academia
  Sinica, P. O. Box 23-141, Taipei 10617, Taiwan}

\altaffiltext{$\dagger$}{Hubble Fellow}

\begin{abstract}

We report observations of three gravitationally lensed supernovae
(SNe) in the Cluster Lensing And Supernova survey with Hubble (CLASH)
Multi-Cycle Treasury program. These objects, SN CLO12Car ($z = 1.28$),
SN CLN12Did ($z = 0.85$), and SN CLA11Tib ($z = 1.14$), are located
behind three different clusters, MACSJ1720.2+3536 ($z = 0.391$),
RXJ1532.9+3021 ($z = 0.345$), and Abell 383 ($z = 0.187$),
respectively.  Each SN was detected in {\it Hubble Space Telescope
  (HST)} optical and infrared images. Based on photometric
classification, we find that SNe CLO12Car and CLN12Did are likely to
be Type Ia supernovae (SNe~Ia), while the classification of SN CLA11Tib is inconclusive. Using multi-color light-curve fits to determine a
standardized SN~Ia luminosity distance, we infer that SN CLO12Car was
$\sim 1.0 \pm 0.2$ mag brighter than field SNe~Ia at a similar
redshift and ascribe this to gravitational lens magnification.
Similarly, SN CLN12Did is $\sim 0.2 \pm 0.2$ mag brighter than field
SNe~Ia. We derive independent estimates of the predicted magnification
from CLASH strong+weak lensing maps of the clusters (in magnitude
units, 2.5\,$\log_{10} \mu$): $0.83 \pm 0.16$ mag for SN CLO12Car,
$0.28 \pm 0.08$ mag for SN CLN12Did, and $0.43 \pm 0.11$ mag for SN
CLA11Tib. The two SNe~Ia provide a new test of the cluster lens model
predictions: we find that the magnifications based on the SN~Ia
brightness and those predicted by the lens maps are consistent.  Our
results herald the promise of future observations of samples of
cluster-lensed SNe~Ia (from the ground or space) to help illuminate
the dark-matter distribution in clusters of galaxies, through the
direct determination of absolute magnifications.

\end{abstract}


\section{Introduction} \label{sec:intro}

Supernovae have proven to be important tools for studying the
Universe. Type Ia supernovae (SNe~Ia) have been instrumental in the
discovery that the expansion of the Universe is accelerating, and
probing the dark energy driving the acceleration \citep[e.g.,][and
  references therein]{Riess1998, Perlmutter1999, Wood-Vasey2007,
  Kessler2009a, Hicken2009, Sullivan2011, Suzuki2012}. Core-collapse
(CC) SNe trace the star-formation history of the Universe
\citep[e.g.,][]{Dahlen2004, Bazin2009, Botticella2012}. Gravitational
lensing of SNe can augment their cosmological utility, in both the
strong and weak lensing regimes. The seminal paper of
\citet{Refsdal1964}, which considered measuring galaxy masses and the
Hubble parameter from lensed image time delays, in fact envisaged SNe
as the background sources.\footnote{\citet{Refsdal1964} also
  presciently concluded with the idea that in addition to lensed SNe,
  ``star-like objects with intense emission both in the radio range
  and optical range'' that ``have been recently discovered'' may also
  be valuable lensed sources, ``giving a possibility of testing the
  different [cosmological] models.''} While no multiply imaged
supernova (SN) has yet been discovered, several applications of lensed
SNe are currently being explored.

Gravitational lensing allows massive galaxy clusters to be used as
cosmic telescopes, increasing the apparent brightness of distant
sources that would otherwise be too faint to detect \citep{Sullivan2000, Kneib2004,
  Stanishev2009}. Ground-based near-infrared (IR) surveys of galaxy
clusters are being used to find lensed SNe, including a core-collapse
SN with redshift $z = 1.7$ \citep[][and references
  therein]{Goobar2009, Amanullah2011}. Lensed SNe~Ia are particularly
useful, as their standardizable luminosities allow for a direct
measurement of the absolute lensing magnification \citep{Kolatt1998}.

The Cluster Lensing And Supernova survey with Hubble
\citep[CLASH;][]{Postman2012} has observed 25 galaxy clusters in 16
{\it Hubble Space Telescope (HST)} broadband filters covering the
near-ultraviolet (UV) to the near-IR with the Advanced Camera for
Surveys (ACS) and the Wide Field Camera 3 (WFC3).  The WFC3 infrared
channel (WFC3/IR) is especially important for finding and studying
high-redshift SNe, efficiently surveying large sky areas for SNe whose
peak flux has redshifted from the optical to the infrared. The SN
discoveries and follow-up observations from CLASH are discussed at
length by \citet{Graur2014}. Here we focus on SNe \emph{behind} CLASH
clusters (in the ``prime'' fields centered on the clusters); these
have not been included in the \citet{Graur2014} ``parallel'' field
sample.

All distant SNe are lensed (magnified, or more often, demagnified) to
some extent by large-scale structure along the line of sight. This
adds a statistical ``nuisance'' in cosmological inferences from SNe~Ia
that can be corrected either assuming an overall magnification
distribution \citep[e.g.,][]{Holz2005, Martel2008} or with convergence estimates
along each SN light of sight using simplified scaling relations
applied to the nearby foreground galaxies observed \citep[][and
  references therein]{Kronborg2010, Jonsson2010, Smith2013}.

Individual SNe that are detectably magnified have been found only
rarely in recent wide-field surveys. A gravitational lens has to be
precisely aligned with the background source to produce large
magnifications, and this is significantly rarer for SNe compared to
background galaxies as sources. Their short-lived durations also make
lensed SNe rarer than even lensed quasars \citep{Oguri2010}. Moreover,
the survey footprints of the Supernova Legacy Survey (SNLS) and the
SDSS-II Supernova Survey did not contain many massive galaxy clusters
to act as lenses for background SNe \citep{Graham2008, Dilday2010}.

\citet{Kronborg2010} inferred that the most magnified SN~Ia in SNLS
had a magnification factor $\mu = 1.27$, corresponding to $\deltam =
2.5\, \log_{10} \mu = 0.26$ mag, calculated from the location and
photometry of foreground galaxies along the line of sight.
\citet{Jonsson2010} came to a similar conclusion, finding that the
SNLS sample did not probe SNe~Ia with $\deltam \gtrsim 0.25$
mag (however, see \citealt{Karpenka2013}, who find little to no lensing of the SNLS sample). Given the typical $\sim 0.2$ mag scatter in standardized SN~Ia
luminosities, a lensing signal at this level cannot be clearly
ascertained for an individual object, although it can be detected
statistically for a larger sample.

Lensing magnification by foreground galaxies is expected to play a
larger role for more distant SNe. The Hubble Deep Field SN~Ia 1997ff
at $z \approx 1.7$ \citep{Gilliland1999, Riess2001} is calculated to
have a $\deltam = 0.34 \pm 0.12$ mag from lens galaxies projected
nearby \citep{Benitez2002}. The uncertainties in the SN light curve
and cosmological model preclude a definitive detection of the expected
magnification signal. Rather, in this case the estimated lensing
magnification has been applied as a correction to the inferred SN
luminosity distance to constrain cosmological parameters, though care
should be taken that this is done self-consistently
\citep{Zitrin2013}.

Supernova properties were used by \citet{Quimby2013} to posit that the
recent SN PS1-10afx at $z = 1.39$ was a SN~Ia magnified by a factor of
30 ($\deltam = 3.72 \pm 0.18$ mag)! However, there is no lens observed
in this system, and the authors conclude that it must be dark.
\citet{Chornock2013} offer an alternate model, that this SN is best
explained as a superluminous core-collapse object, without significant
magnification.

Ideally, we would like to confirm the detection of gravitational
lensing magnification of a SN by independently checking the
magnifications derived from the SN brightness and a lens-model
prediction for consistency. In this paper, we discuss three
gravitationally lensed SNe (SN CLO12Car, SN CLN12Did, and SN CLA11Tib)
behind CLASH clusters. We present photometric observations of the SNe
and spectroscopy of their host galaxies in \S 2. In \S 3 we discuss
classification of the SNe and model light-curve fits. The
gravitational magnifications from lens models are derived in \S 4, and
we confront the models against the observations and discuss the
results in \S 5. The following cosmological parameters are adopted
throughout the paper \citep{Sullivan2011}: \Hn, $\Mat$ = 0.27, $\Lam$ = 0.73, and $\Cur$ = 0.

\section{Observations} \label{sec:obs}
 
\subsection{SN CLO12Car} \label{sec:obs-car}

\begin{figure}
\includegraphics[angle=0,width=3.4in]{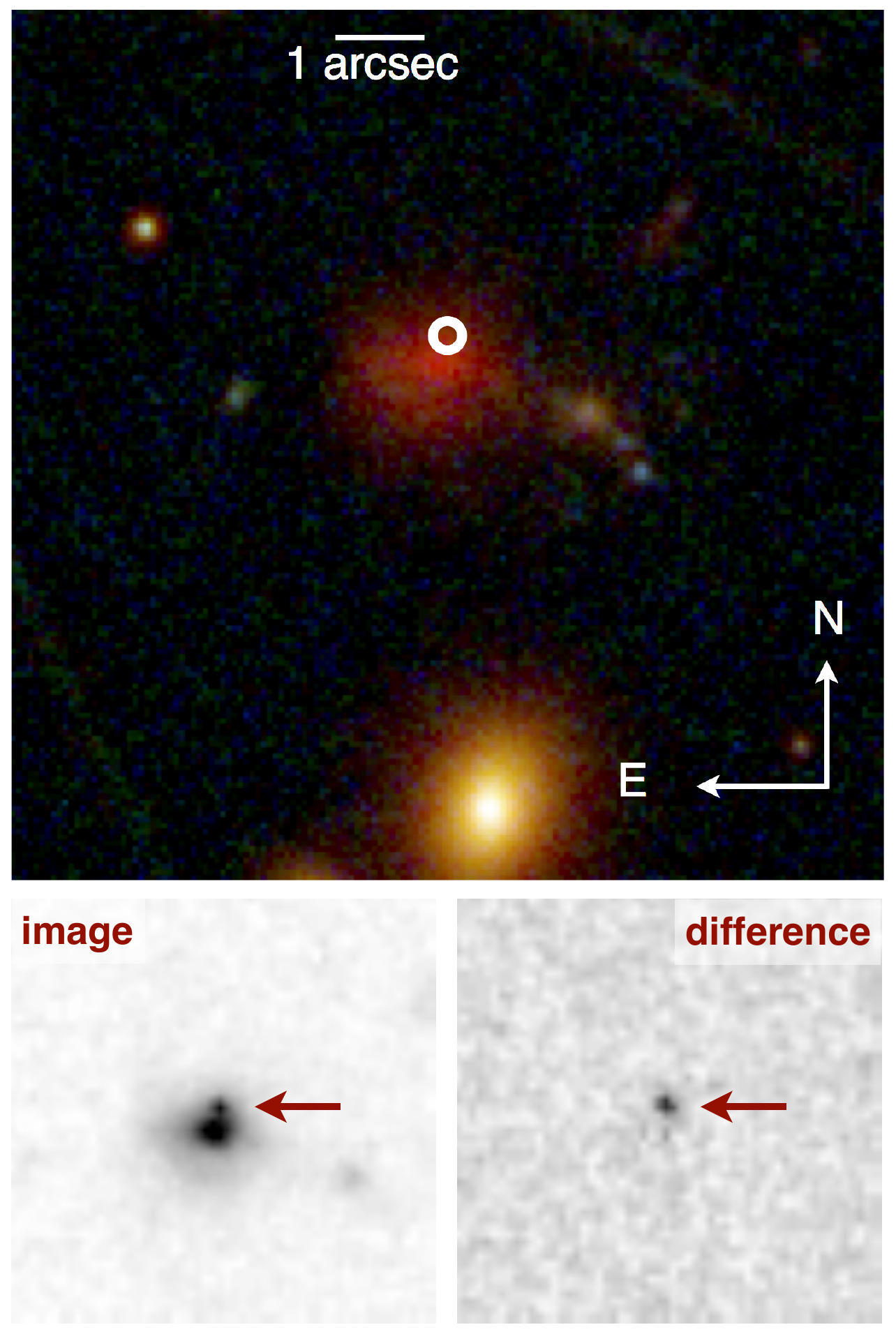}
\caption{\emph{Top:} CLASH \emph{HST} composite false-color image of SN
  CLO12Car.  In this image, the red channel is an average of the
  WFC3/IR F105W, F110W, F125W, F140W, and F160W data. The
  green channel combines ACS F606W, F625W, F775W, 
  F814W, and F850LP observations, and the blue channel is from ACS F435W 
  and F475W data. The location of the SN is marked with a white
  circle. \emph{Bottom:} The left panel shows a $5''$ square cutout
  of the WFC3/IR F160W image from 2012 June 17, with SN CLO12Car marked
  by the arrow; the right panel shows the difference image after
  template subtraction. \label{fig:car-image}}
\end{figure}

SN CLO12Car was discovered in the CLASH images of
MACSJ1720.2+3536 \citep{Ebeling2010}. The SN was detected in June 2012 with both ACS and
WFC3/IR. Figure \ref{fig:car-image} shows the location of the SN
($\alpha = 17^{\rm h}20^{\rm m}21.03^{\rm s}$, $\delta =
+35^{\circ}36'40\farcs9$, J2000), 0.35\arcsec\ from the center of its
host galaxy.  The host is 
relatively bright in the IR (F160W = 21.50 $\pm$ 0.01 AB mag), but is
significantly fainter in the optical images (F606W = 24.21 $\pm$ 0.06
AB mag).

We present aperture photometry for SN CLO12Car in
Table~\ref{tab:car-phot}. These data include publicly available
follow-up observations collected in July 2012 by {\it HST} program
GO-12360 (PI S.~Perlmutter). The flux from SN CLO12Car was measured in
difference images, using templates constructed with CLASH observations
from March and April 2012, more than 30 days before the first
detection.  For the ACS images, we used a fixed 4-pixel ($0\farcs20$)
radius aperture, whereas for WFC3/IR, the photometry was measured in a
3-pixel radius ($0\farcs27$) aperture.

\begin{figure*}
\includegraphics[width=\textwidth]{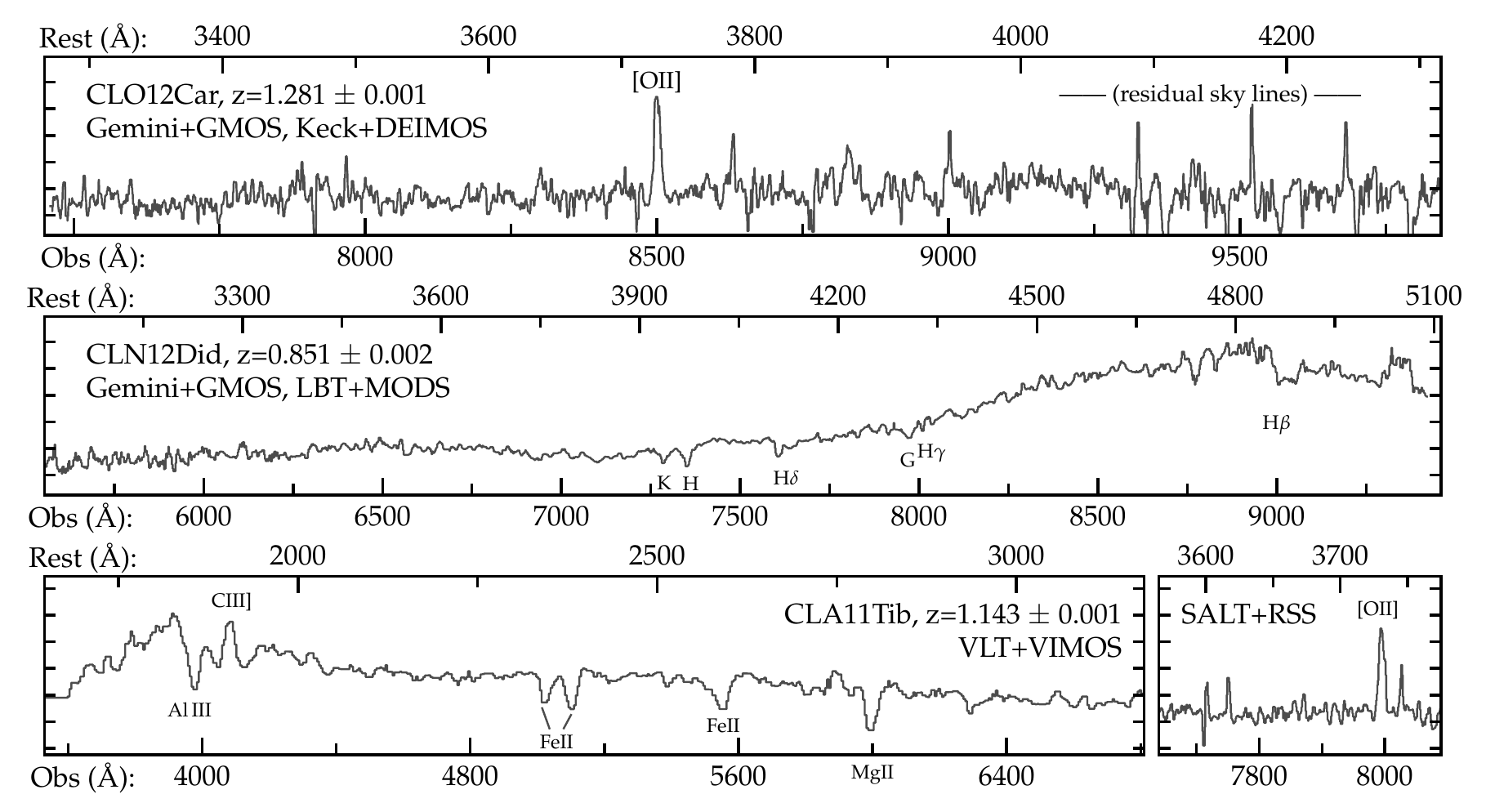}
\caption{ \label{fig:HostSpec} Spectra of the host galaxies of the
  three lensed SNe.  Each panel shows observed wavelength along the
  bottom axis and rest wavelength on top. For display purposes, all
  spectra have been smoothed with a running 3-pixel median.
  \emph{Top}: The host galaxy of SN CLO12Car, at $z=1.281 \pm 0.001$.
  The plot shows the mean spectrum combining observations with
  Gemini+GMOS and Keck+DEIMOS.  The single emission line at 8500
  \AA\ is identified as the [\ion{O}{2}] $\lambda\lambda 3726, 3728$
  doublet based on the photo-$z$ and asymmetry of the line
  profile. \emph{Middle}: The CLN12Did host galaxy, combining spectra
  from Gemini+GMOS and LBT+MODS. We derive $z=0.851 \pm 0.002$ from
  cross-correlation with galaxy templates and identification of Ca and
  H absorption features, as indicated.  \emph{Bottom}: The left panel
  shows the spectrum of the SN CLA11Tib host galaxy, observed with
  VLT+VIMOS. We derive $z=1.143 \pm 0.001$ from multiple absorption
  features and \ion{C}{3}] $\lambda$1909 emission. This confirms the
        [\ion{O}{2}] emission-line redshift from the SALT+RSS spectrum
        shown on the right.}
\end{figure*}

We used the DEep Imaging Multi-Object Spectrograph
\citep[DEIMOS;][]{Faber2003} on the Keck-II 10~m telescope to obtain a
$\sim 2$ hr spectrum of the host galaxy of SN CLO12Car ($2 \times
1800$ s, $1 \times 1700$ s, and $1 \times 1500$ s), with $\sim 4$
\AA\ resolution, covering the wavelength range 5000--9000 \AA.  The
spectrum showed evidence of a weak emission line located between two
night-sky lines near 8506 \AA. For confirmation, we obtained a 2.5 hr
($5 \times 1800$ s) Gemini North Multi-Object Spectrograph
\citep[GMOS;][]{Hook2004} nod-and-shuffle spectrum of the host,
spanning 7460--9570 \AA\ with a resolution of $\sim 2.1$ \AA. The GMOS
spectrum also shows evidence for the faint line (marginally resolved
as a doublet), which we identify as [\ion{O}{2}] $\lambda\lambda 3726,
3728$ emission at $z = 1.281 \pm 0.001$.  A composite of the DEIMOS and
GMOS spectra is shown in the top panel of Figure~\ref{fig:HostSpec}.
The spectroscopic redshift for the CLO12Car host galaxy lies near the
peak of its photometric redshift (photo-$z$) probability density
function (PDF), but the photo-$z$ PDF is broad, only limiting the
redshift to $0.9 \lesssim z \lesssim 1.9$. A full description of the
CLASH photo-$z$ estimation methods is given by \citet{Jouvel2013} and
Molino et al.~(in prep.). At $z=1.281$, a separation of 
0.35\arcsec\ from SN to host-galaxy center corresponds to a projected physical 
separation of $\sim 3$ kpc (modulo lensing magnification).

We also analyzed the follow-up \emph{HST} WFC3/IR G102 and G141 grism
\citep{Dressel2012} data for SN CLO12Car and its host galaxy (taken
while the SN was still visible; program GO-12360, PI
S.~Perlmutter). The G102 spectrum has a dispersion of 25
\AA\ pixel$^{-1}$, covering 8000--11,500 \AA, while the G141 spectrum
has a dispersion of 47 \AA\ pixel$^{-1}$, covering 11,000--17,000 \AA.
The exposure time with each grating was about 1.3 hr, but this yielded
only a weak signal, and we were unable to extract a clear SN
spectrum. In these slitless grism observations, spectral features are
effectively convolved with the spatial profile of the source, making
sharp emission lines broader and less pronounced.  There is an
emission line in the G141 grism spectrum at 14,963 \AA, consistent
with H$\alpha$ $\lambda$6563 at $z \approx 1.28$, but this is only a
marginal detection given the low resolution of the data.

\subsection{SN CLN12Did} \label{sec:obs-did}

\begin{figure}
\includegraphics[angle=0,width=3.4in]{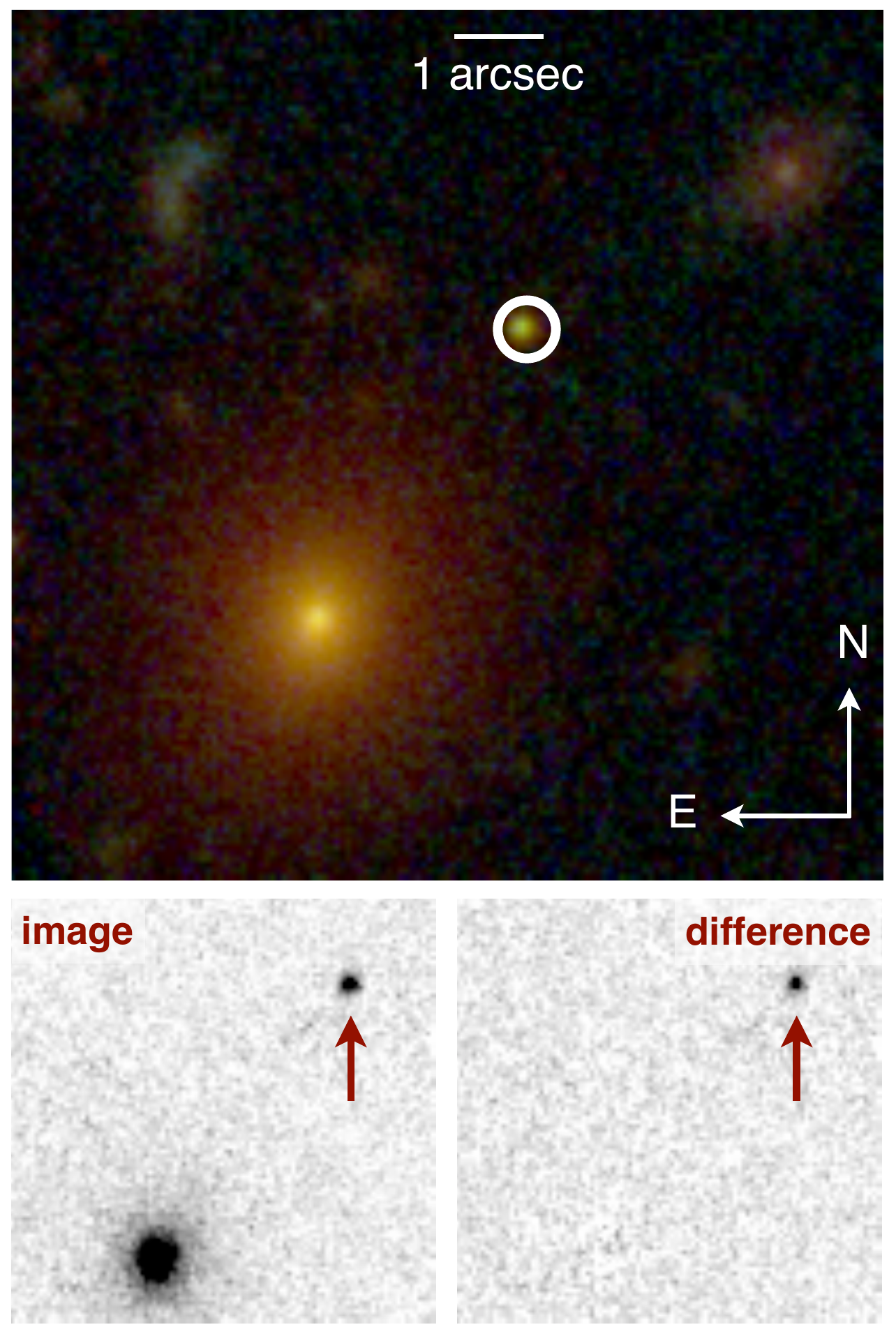}
\caption{\emph{Top: HST} composite false-color image of SN CLN12Did and
  surroundings. The red channel is an average of WFC3/IR F105W, F110W,
  F140W, and F160W data from the CLASH program, the green channel
  combines the ACS F606W, F625W, F775W, F814W, and F850LP
  observations, and the blue channel is from ACS F435W and F475W
  data. The SN position is marked with a white circle, in the
  outskirts of its early-type host galaxy to the
  southeast. \emph{Bottom:} The left panel shows a $5''$ square cutout
  of the ACS F850LP image from 2012 March 3, with SN CLN12Did marked
  by the arrow; the right panel shows the difference image after
  template subtraction. \label{fig:did-image}}
\end{figure}

SN CLN12Did was discovered in early 2012 in CLASH ACS and WFC3/IR
images of RXJ1532.9+3021 \citep{Ebeling1998}. Figure \ref{fig:did-image} shows the
position of the SN (located at $\alpha = 15^{\rm h}32^{\rm
  m}59.25^{\rm s}$, $\delta = +30^{\circ}21'42\farcs8$, J2000), and we
present aperture photometry of the SN in Table \ref{tab:did-phot}. The
SN was present in the first epoch of observations in each of the ACS
and WFC3/IR filters, so we did not have a SN-free template for
subtraction. Nonetheless, the SN is separated by $\sim 4''$ from its
putative host galaxy, so we expect negligible host-galaxy
contamination in the SN photometry.

Ground-based Sloan Digital Sky Survey (SDSS) DR7 images
\citep{Abazajian2009} of the CLN12Did host, along with the {\it HST}
data, show the CLN12Did host to be an early-type, likely elliptical,
galaxy. The CLASH data yield photometry for the host galaxy of F606W =
22.69 $\pm$ 0.03 AB mag and F160W = 19.434 $\pm$ 0.003 AB mag. We
obtained a $\sim 5$ hr spectrum ($6 \times 2700$ s and $1 \times 2500$
s) of the host galaxy using the Multi-Object Double Spectrographs
\citep[MODS;][]{Pogge2010} on the Large Binocular Telescope (LBT),
covering 6300--9000 \AA\ with 15--40 \AA\ resolution. We also used
Gemini North GMOS to obtain a 4.5 hr ($9 \times 1800$ s) spectrum of
the galaxy, covering 5220--9420 \AA, with a resolution of $\sim 6$
\AA. A composite spectrum from the Gemini and LBT observations is
shown in the middle panel of Figure~\ref{fig:HostSpec}.  We derived
respective redshifts of 0.851 $\pm$ 0.001 and 0.852 $\pm$ 0.002 from
these data, via cross-correlation with absorption-line templates.  This
redshift is consistent with the photo-$z$ derived for the galaxy ($z_{phot} = 0.92^{+0.05}_{-0.07}$). We
adopt $z = 0.851 \pm 0.001$ for SN CLN12Did.  At this redshift, the $4''$ separation from SN to host-galaxy center 
corresponds to a projected physical separation of $\sim 30$ kpc (modulo lensing magnification).

\subsection{SN CLA11Tib} \label{sec:obs-tib}

\begin{figure}
\includegraphics[angle=0,width=3.4in]{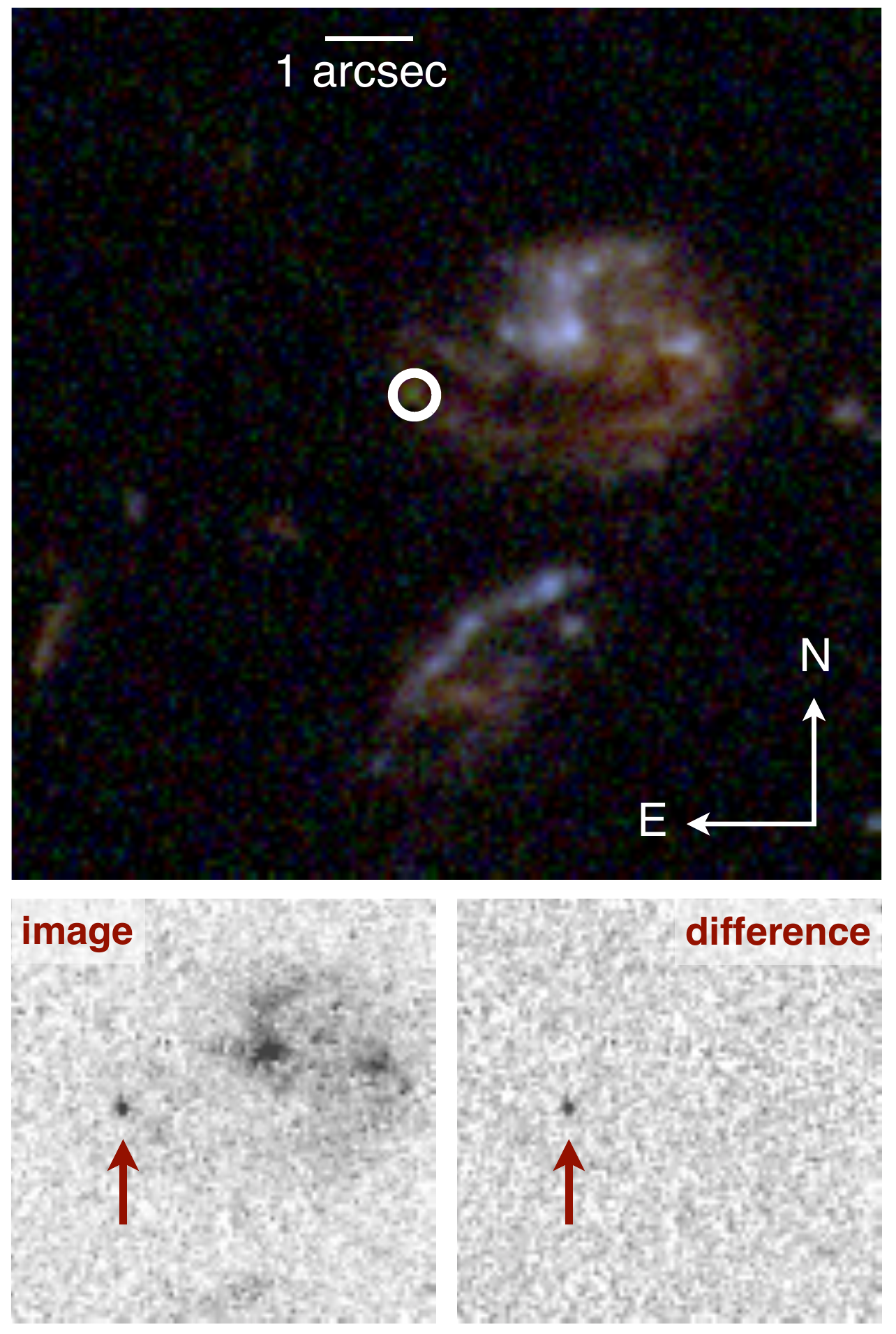}
\caption{\emph{Top: HST} composite false-color image of SN CLA11Tib
  from CLASH data. The red channel is an average of WFC3/IR F105W,
  F110W, F125W, F140W, and F160W data.  The green channel combines ACS
  F606W, F625W, F775W, F814W, and F850LP observations, and the blue
  channel is from ACS F435W and F475W data. The location of the SN is
  marked with a white circle. \emph{Bottom:} The left panel shows a
  $5''$ square cutout of the ACS F850LP image taken on UT 2011 Jan.~4,
  with SN CLA11Tib indicated by the arrow; the right panel shows the
  difference image after template subtraction. \label{fig:tib-image} }
\end{figure}

SN CLA11Tib was discovered in CLASH ACS and WFC3/IR images of Abell
383 \citep{Abell1989} in January 2011. Figure \ref{fig:tib-image} shows the SN in a
color image (located at $\alpha = 02^{\rm h}48^{\rm m}01.27^{\rm s}$,
$\delta = -03^{\circ}33'16\farcs9$, J2000). Aperture photometry for the ACS observations of SN CLA11Tib is presented in Table \ref{tab:tib-phot}.  The WFC3/IR
observations of the SN were taken as part of the {\it HST} follow-up
program GO-12360 (PI S.~Perlmutter). The ACS photometry is from
difference images with host-galaxy subtraction, but no template images
are available for the WFC3/IR observations (F105W, F125W, and F160W),
so the reported IR photometry is based on the direct images. As the SN is located in a spiral arm (see Figure \ref{fig:tib-image}), the host galaxy could contaminate aperture photometry for these observations. For this reason we report  point-spread function (PSF) fitting photometry of the SN in the WFC3/IR observations from Dolphot \citep[a modified version of HSTphot;][]{Dolphin2000}. Photometry of the host
galaxy gives F606W = 22.88 $\pm$ 0.01 AB mag and F160W = 21.51 $\pm$
0.01 AB mag.

We obtained Southern African Large Telescope (SALT) + Robert Stobie
Spectrograph \citep[RSS;][]{Nordsieck2012,Crawford2010} spectroscopy
of the host of SN CLA11Tib with a 1200 s exposure using the PG0900
grating (range 6300--9300 \AA, resolution 6 \AA), and detect a strong
emission line (but close to a night-sky line) that we identify as
[\ion{O}{2}] at $z = 1.144 \pm 0.001$. We confirmed this
identification with the VIsible MultiObject Spectrograph
\citep[VIMOS;][]{LeFevre2003} on the Very Large Telescope (VLT)
observations in a $\sim 4$ hr spectrum of the host galaxy (range
3900--6700 \AA\ with 28 \AA\ resolution). The spectrum shows
\ion{Mg}{2} and \ion{Fe}{2} absorption, and \ion{C}{3}] $\lambda$1909
  emission, all at a consistent redshift of $1.143 \pm 0.001$, which
  we adopt for SN CLA11Tib. This redshift is consistent with the 
  photo-$z$ derived for the galaxy ($z_{phot} = 1.05 \pm 0.10$).  At this redshift, SN 
  CLA11Tib is projected $\sim 25$ kpc from the center of its host galaxy (uncorrected for lensing magnification).

\section{SN Classification and Light-Curve Fits} \label{sec:class}

Classification of high-redshift ($z \gtrsim 1$) SNe is challenging
\citep[e.g.,][]{Rodney2012, Rubin2013, Jones2013}. For example, a
typical SN~Ia has a peak luminosity in the rest-frame $B$ band of $M_B
\approx -19.4$ mag. At $z = 1.5$ this becomes a $Y$-band apparent
magnitude of $\sim 25.8$.  Key spectral absorption features at
rest-frame optical wavelengths (such as the \ion{Si}{2} feature
observed in SN~Ia at $\sim 6150$ \AA) are redshifted into the IR,
making them inaccessible to most ground-based spectrographs.  With
these constraints, definitive spectroscopic classification of high-$z$
SNe requires prohibitively large investments of \emph{HST} time
\citep{Jones2013}.

For this reason, we rely on photometric classifications for these
three lensed SNe.  We utilize a Bayesian photometric SN classifier
called STARDUST: Supernova Taxonomy And Redshift Determination Using
SNANA Templates.  An early version of this classifier was first
introduced by \citet{Jones2013}, and the complete version used for
this work is presented in more detail by \citet{Graur2014} and Rodney
et al.~(in prep.). In brief, observations are compared against
simulated SNe~Ia and CC~SNe, generated using version 10.27k of the
SuperNova ANAlysis software
package\footnote{http://sdssdp62.fnal.gov/sdsssn/SNANA-PUBLIC/}
\citep[SNANA;][]{Kessler2009b}.  For SNe~Ia the STARDUST classifier
uses the SALT2 light-curve model \citep{Guy2010} extended to cover the rest-frame near-UV and IR wavelengths (see below), whereas for CC~SNe,
the classifier draws on the SNANA library of 16 SN~Ib/c and 26 SN~II
templates, to be described in further detail by \citet{Sako2014}. An important capability of STARDUST that we employ for this
work is to allow an achromatic flux scaling factor for every model
realization. This means that our STARDUST classifications rely only on
the observed SN colors and light curve shape, and do not depend on the intrinsic SN
luminosities or our choice of cosmological model. This is especially
important for fitting lensed SNe, as it allows us to classify the SNe
independent of the lensing magnification.

For additional verification of our photometric SN classifications, we
turned to the Photometric SuperNova IDentification software
\citep[PSNID;][]{Sako2008, Sako2011} to classify the three SNe. PSNID
is similar to the STARDUST classifier, as both use SALT2 to simulate
SNe~Ia and rely on the same 16 SN~Ib/c and 26 SN~II templates to
simulate CC~SNe. However, PSNID has the advantage that it has been
thoroughly tested, obtaining the highest figure of merit in the SN
Photometric Classification Challenge \citep{Kessler2010}, and it is
already publicly available as a component of SNANA.  A downside of
PSNID for this work is that it uses absolute magnitude priors in its
classifications, which is not ideal for analyzing lensed SNe.

For SNe~Ia, we also employed a modified version of the \citet{Guy2010}
SALT2 model to fit light curves and derive standardized luminosity
distances. The version of SALT2 presented by \citet{Guy2010} has been
robustly tested, and was used in SNLS cosmology papers
\citep{Sullivan2011,Conley2011}. To accommodate the wide wavelength
range of our data, we extended the SALT2 model to cover rest-frame
near-UV and IR wavelengths (by default, the SALT2 fitter can only fit
rest wavelengths of 2800--7000 \AA). For all three SNe, this extended SALT2 model produces similar results (within 1$\sigma$) to those of the original \citet{Guy2010} model. In deriving distances, we
converted our SALT2 parameters representing light-curve shape and
color excess ($x1$ and $c$) to their respective SiFTO
\citep{Conley2008} parameters ($s$ and $C$), using the conversion
equations from \citet{Guy2010}. This was done in order to adopt the
constants $M$, $\alpha$, and $\beta$, from \citet{Sullivan2011} and
also used by \citet{Jones2013}. Specifically, the distance modulus was
calculated using the following equation:\footnote{We use the notation
  ``$\dm$'' for distance modulus, rather than the conventional $\mu$,
  to avoid confusion with the lensing magnification $\mu$. As usual,
  it is defined as $\dm = 5\, \log_{10} d_L + 25$, where $d_L$ is the
  luminosity distance in Mpc.}
\begin{equation}
  \dm_{\rm SALT2} = m_{B}^{*} - M + \alpha(s-1) - \beta C,
\end{equation}
with $M = -19.12 \pm 0.03$, $\alpha = 1.367 \pm 0.086$, and $\beta =
3.179 \pm 0.101$. To match the (for our purposes, arbitrary)
normalization of the SALT2 fitter used by \citet{Guy2010} to our
results from SNANA, we applied an offset of 0.27 mag to the value of
$m_{B}^{*}$ reported by SNANA.

We also used the \citet{Jha2007} version of MLCS2k2 included in SNANA
to fit the SNe~Ia. To check for consistency, we compared the best-fit
light curve shape parameter $\Delta$ from MLCS2k2 to $x1$ from SALT2
using the relationship in Appendix G given by \citet{Kessler2009a}. To
compare the SALT2 color parameter $c$ with MLCS2k2 $A_V$, we employ a
linear relationship derived from Figure 42 of \citet{Kessler2009a},
specifically
\begin{equation}
c = (0.464 \pm 0.021)A_V - (0.121 \pm 0.014).
\end{equation}

Finally, to ensure a consistent zeropoint between the SALT2 and
MLCS2k2 distance moduli, we applied both methods to a sample of SDSS
Supernova Survey SNe~Ia from \citet{Holtzman2008, Kessler2009a}. We
added a zeropoint correction of 0.20 mag to the distance moduli
reported by SNANA in the MLCS2k2 fit ($\dmm$), to put them on the same
scale as the SALT2 distances ($\dms$).

Although we report magnitudes in Tables~\ref{tab:car-phot}, \ref{tab:did-phot}, and \ref{tab:tib-phot}, both the classification and SN~Ia light curve fitting were done in flux space. The upper limits for each SN were included in the fits with the measured flux and 1$\sigma$ uncertainties. 

\subsection{SN CLO12Car} \label{sec:class-car}

SN CLO12Car was located in a red disky galaxy, in which both SNe~Ia
and CC~SNe could be expected. The STARDUST classifier prefers a SN~Ia
fit with probability ($\PIa$) of 0.91. The
probabilities for SN~Ib/c ($\PIbc$) and SN~II ($\PII$) models are
$~10^{-8}$ and 0.09, respectively. The best-fit SN~Ia, SN~Ib/c,
and SN~II models had $\chi^2/\nu$ ($\nu$ = 11) of 1.12, 4.56, and 1.76,
respectively. PSNID also favors a SN~Ia fit at a high confidence, with
$\PIa = 0.99$ and a best-fit $\chi^2/\nu = 1.06\ (\nu = 11)$\footnote{PSNID uses a luminosity prior, as discussed above. Correcting the light curve for lensing, assuming reasonable values of magnification between 1 mag of demagnification and 3 mag of magnification, changes the $\PIa$ value slightly (ranging from 0.87 to 0.99).}. We thus
conclude that SN CLO12Car was a SN~Ia. 

\begin{figure}
\begin{center}
\includegraphics[width=0.45\textwidth]{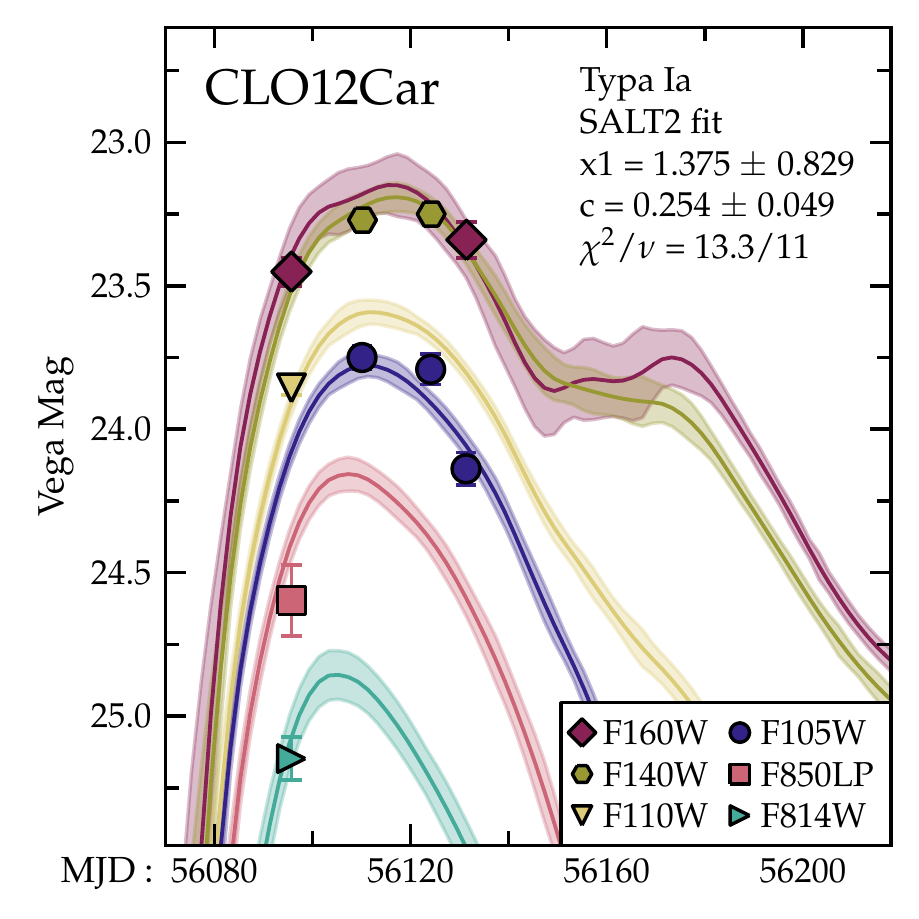}
\caption{ \label{fig:car-lc} SALT2 light-curve fit of SN CLO12Car. The
  light-curve parameters $x1$ and $c$ are typical of normal SNe~Ia,
  and the data are well matched by the model, with $\chi^2/\nu$ =
  13.3/11.}
\end{center}
\end{figure}

Both MLCS2k2 and SALT2 produce good light-curve fits for SN
CLO12Car. We show the SALT2 light-curve fit along with the 1$\sigma$
model errors in Figure \ref{fig:car-lc}. The SALT2 fit had a
$\chi^2/\nu$ = 13.3/11 = 1.21, and the fit parameters ($x1 = 1.375 \pm
0.829$ and $c = 0.254 \pm 0.049$) are in the range of normal SNe~Ia
\citep{Guy2010}. From these parameters, we derive a distance modulus
of $\dms = 43.83 \pm 0.24$ mag for the SN. 

The MLCS2k2 light-curve fit had a slightly worse, but still
acceptable, goodness-of-fit ($\chi^2/\nu$ = 18.1/11 = 1.64). The model
parameters were $\Delta = 0.007 \pm 0.314$ and $A_V = 0.635 \pm 0.190$
mag. The light curve shape parameters from the two fitters agree at $<
1\sigma$; specifically, the SALT2 $x1$ value corresponds to $\Delta =
-0.380 \pm 0.160$. The best-fit MLCS2k2 $A_V$ corresponds to $c =
0.174 \pm 0.090$, which is also $< 1\sigma$ from the best-fit SALT2
color. Moreover, we find that $\dmm = 43.71 \pm 0.16$ mag (after
applying the zeropoint correction), again consistent with the inferred
$\dms$ at $< 1\sigma$.

For comparison, at $z = 1.281$, the distance modulus expected from
$\Lambda$CDM with our adopted cosmological parameters is $\dmLCDM =
44.76$ mag, significantly higher than what is measured from the SN
light curve. We discuss the implications of this finding in \S
\ref{sec:lensing}.

\subsection{SN CLN12Did} \label{sec:class-did}

SN CLN12Did was located in an elliptical galaxy, already indicating it
was most likely a SN~Ia \citep{Cappellaro1999, Mannucci2005,
  Foley2013}.  The STARDUST classifier prefers a SN~Ia fit at high
confidence ($>5\sigma$). PSNID also favors a SN~Ia fit with high
significance ($\PIa > 0.9999$), with best-fit $\chi^2/\nu = 0.5\ (\nu
= 20)$. These consistent results yield the strong conclusion that SN
CLN12Did was a SN~Ia.

\begin{figure*}
\begin{center}
\includegraphics[width=\textwidth]{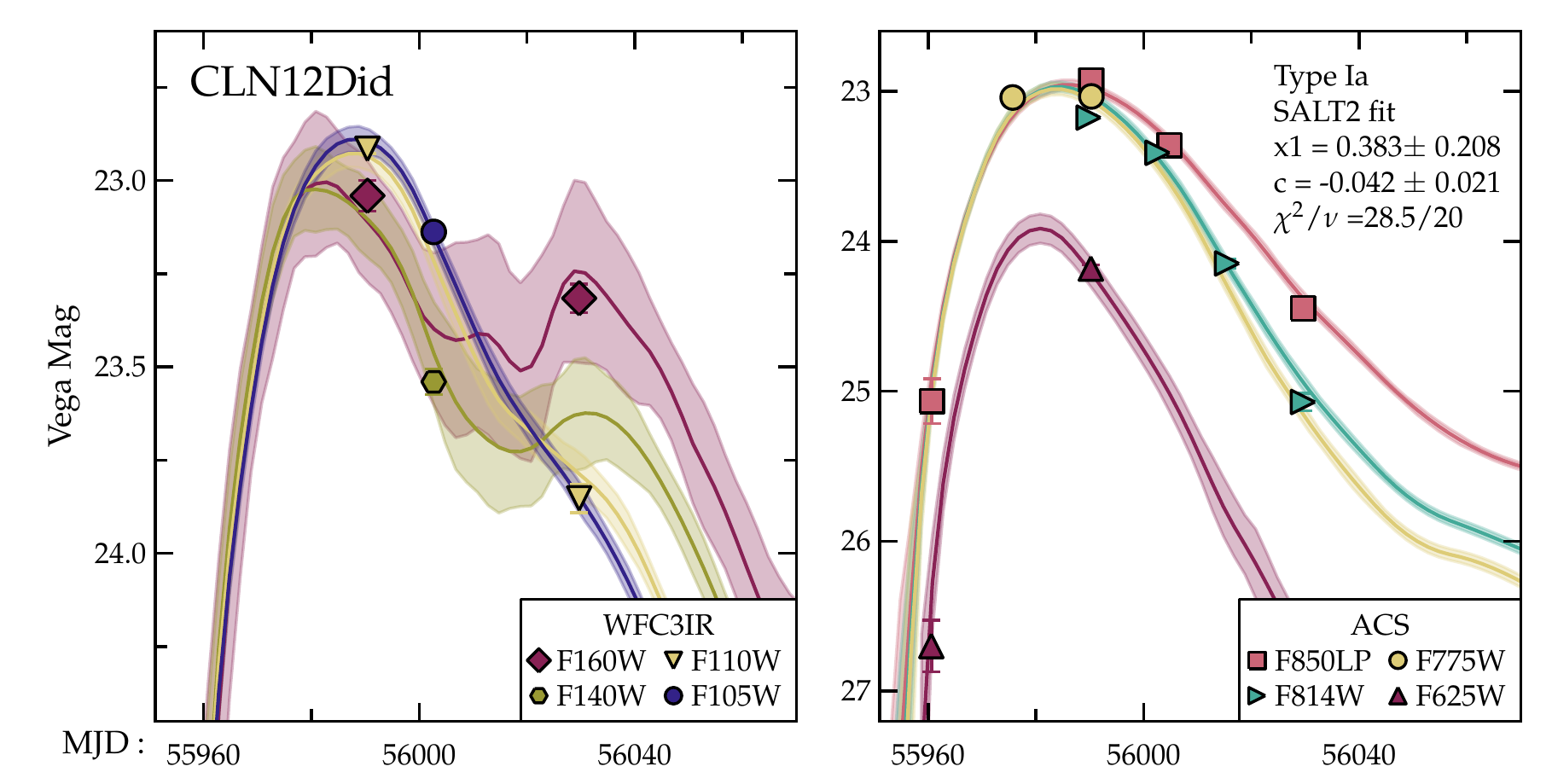}
\caption{ \label{fig:did-lc} SALT2 light-curve fit of SN CLN12Did.
 Upper limits on F435W and F475W nondetections are excluded from the plot for
clarity, but all of the observations listed in Table~\ref{tab:did-phot} are used in the
light-curve fit. This model is a reasonably good match to the data, yielding
  $\chi^2/\nu = 1.4$.  The best-fit light-curve parameters are typical
  for a normal SN~Ia.}
\end{center}
\end{figure*}

We show the SALT2 SN~Ia light-curve fit for SN CLN12Did in Figure
\ref{fig:did-lc}. Similar to Figure \ref{fig:car-lc}, we display the
light-curve points and the model curves with 1$\sigma$ errors.  Upper
limits on F435W and F475W nondetections are excluded from the plot for
clarity, but all of the observations listed in Table~\ref{tab:did-phot} are used in the
light-curve fit. The SALT2 parameters fall within the normal SN~Ia
range, with $x1 = 0.383 \pm 0.208$, $c = -0.042 \pm 0.021$, and
$\chi^2/\nu = 28.5/20 = 1.42$. From the light-curve parameters, we
derive $\dms = 43.33 \pm 0.15$ mag. 

MLCS2k2 produces a slightly better light-curve fit, with $\chi^2/\nu =
23.6/20 = 1.18$. The best-fit parameters are $A_V = 0.042 \pm 0.059$
mag and $\Delta = 0.055 \pm 0.098$. Using the conversion equations,
the MLCS2k2 fit $A_V$ corresponds to SALT2 $c = -0.102 \pm 0.031$, and
the SALT2 $x1$ predicts MLCS2k2 $\Delta = -0.212 \pm 0.040$. This is
good agreement in the color parameter, but a slight ($\sim 2.5\sigma$)
discrepancy in light-curve shape. The inferred distance is $\dmm =
43.28 \pm 0.09$ mag, consistent with the SALT2 result.  For
comparison, at $z=0.851$, the cosmological parameters yield $\dmLCDM =
43.65$ mag.

\subsection{SN CLA11Tib} \label{sec:class-tib}

SN CLN12Tib is located in a spiral arm of its blue, late-type host
galaxy, an environment in which all SN types can be found
\citep{Cappellaro1999, Mannucci2005, Li2011, Foley2013}. The STARDUST
classifier prefers a CC fit for this SN, returning probabilities $\PII
= 0.41$ and $\PIbc = 0.52$, compared to the SN~Ia fit probability $\PIa=
0.07$.

\begin{figure*}
\begin{center}
\includegraphics[angle=0,width=\textwidth]{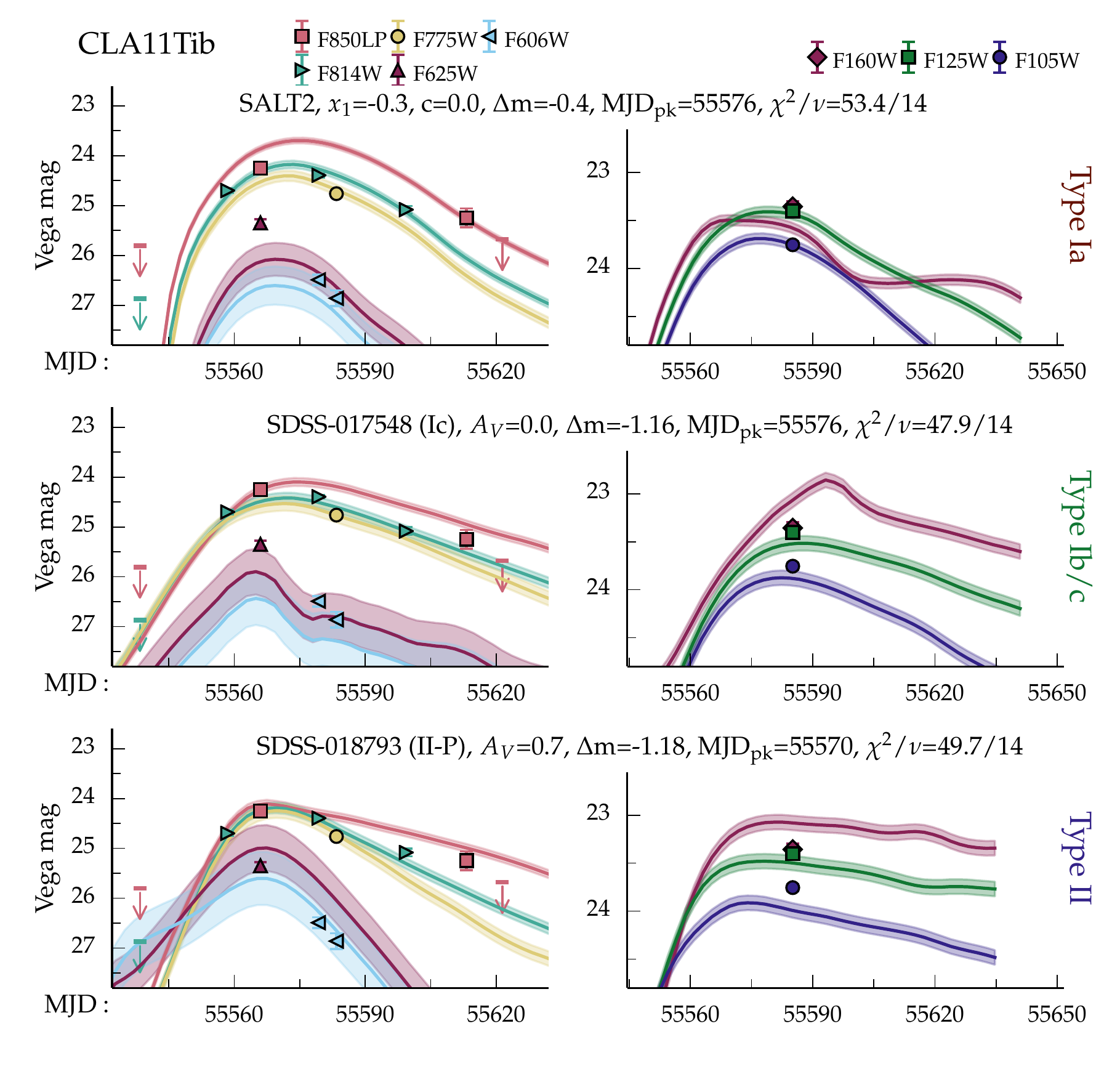}
\caption{\label{fig:tib-lc} STARDUST light-curve fits to CLA11Tib for
  the three primary SN subclasses.  The top row shows the maximum
  likelihood fit from a SN~Ia (SALT2) model, the middle row displays
  the best-fit SN~Ib/c model, and the bottom row shows the preferred
  SN~II model. Optical data from ACS appear in the left column, while
  WFC3/IR observations are on the right. The CC~SN models are labeled
  by the best-fit template SN, and all three models show the best-fit
  parameters. The downward pointing arrows indicate 3$\sigma$ upper
  limits on the SN flux. The $\Delta m$ term designates the offset
  (mag) needed to make the template match the data, discussed further
  in \S \ref{sec:lensing-tib}.}
\end{center}
\end{figure*}

\begin{figure*}
\begin{center}
\includegraphics[width=5.5in]{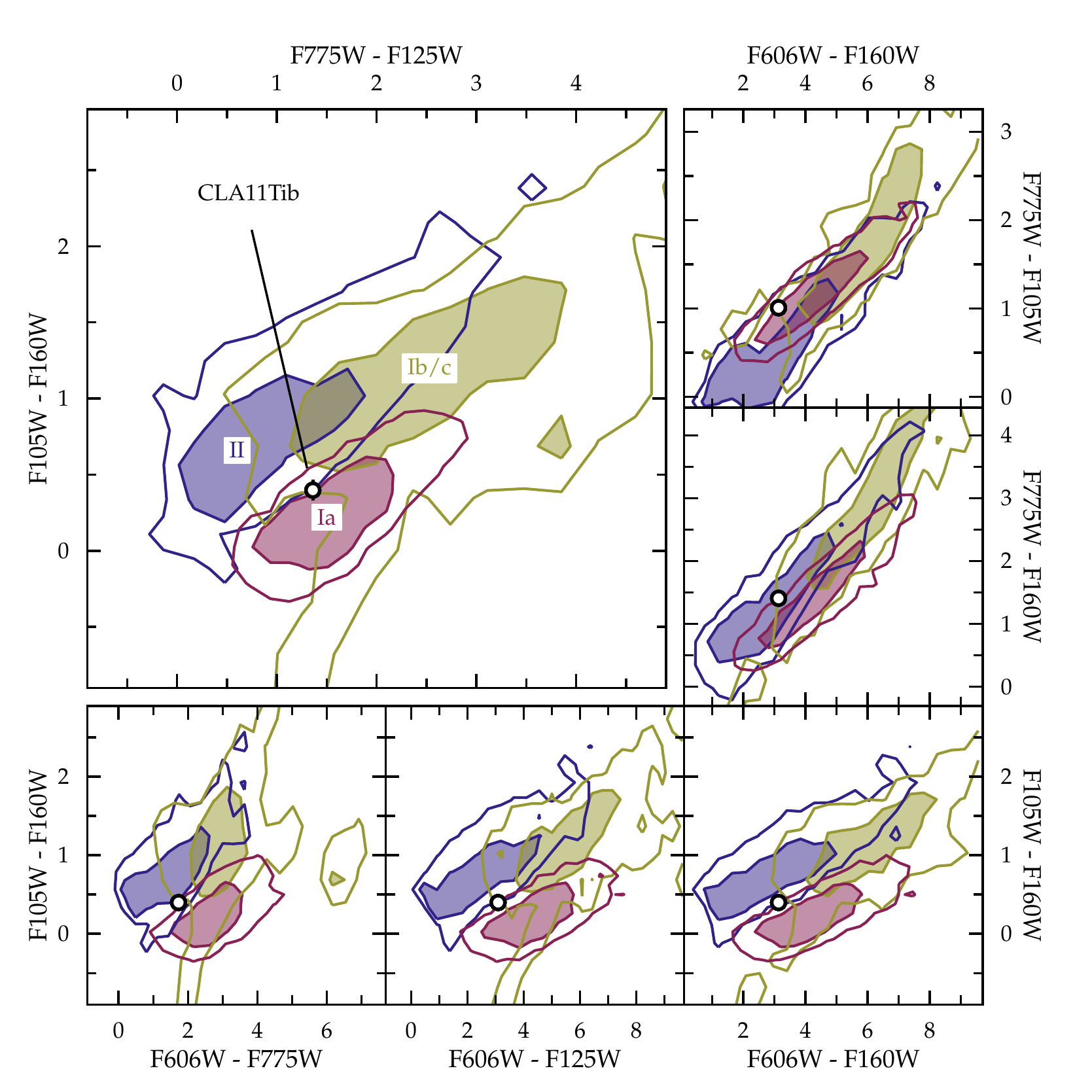}
\caption{Color-color diagrams for SN CLA11Tib photometry from MJD
  55583--55585, illustrating the difficulty in photometric
  classification of this object. We utilize color-based classification
  so as to be unaffected by achromatic gravitational lensing. The data
  favor a core-collapse origin for this SN (either SN~Ib/c or
  SN~II). The contours shown are derived from SNANA Monte Carlo
  simulations with 5000 simulated SNe for each of the 3 subclasses (so
  they do not reflect the relative frequency across classes). Shaded
  regions enclose 68\% of each population and solid lines enclose
  95\%. \label{fig:tib-colorcolor}}
\end{center}
\end{figure*}

We show the light-curve fit for SN CLA11Tib in Figure
\ref{fig:tib-lc}, with each row giving the maximum-likelihood fit for
a single SN subclass. The $\chi^2/\nu$ ($\nu$ = 14) values for the
best-fit SN~Ia, SN~Ib/c, and SN~II models are 3.81, 3.42, and 3.55,
respectively. None of the three SN types are a good fit to the light-curve; moreover, none of the templates provide a good fit to all the WFC3/IR data. There are only 42 CC~SNe templates, and we do not expect them to represent the full range of SNe~II and SNe~Ibc.
The SN~II template that provides the best match to the
observations is SDSS-018793 (SN 2007og), a SN~IIP at $z=0.20$
\citep{Sako2014}.  For the SNe~Ib/c, the best match comes from SN
SDSS-017548 (SN 2007ms), a SN~Ic at $z = 0.0393$ \citep{Ostman2011}.
The sole \emph{HST} WFC3/IR F125W observation of SN
CLA11Tib is close to the peak for both the SN~IIP and SN~Ic models; at
$z = 1.143$, the F125W filter corresponds to approximately the rest
$V$ band. To further highlight the difficulty in photometrically
classifying SN~CLA11Tib, in Figure \ref{fig:tib-colorcolor} we show
color-color diagrams from the ACS and WFC3/IR photometry taken during
2011 Jan.~22--24, where the SN straddles the typical colors of SNe~Ia,
Ib/c, and II, though favoring CC~SN models. The full STARDUST
classification uses these types of constraints with all available
epochs.

Our PSNID results disagree with those from the STARDUST
classifier. PSNID favors a SN~Ia fit for SN CLA11Tib at greater than
4$\sigma$. However, the best-fit PSNID SN~Ia model does not match the
light curve well ($\chi^2/\nu$ = 2.83, with $\nu$ = 14). To check these
results, we fit SN CLA11Tib with SALT2 and MLCS2k2 directly using
SNANA. The best SALT2 fit has normal light-curve parameters, with $x1
= 0.114 \pm 0.486$ and $c = 0.048 \pm 0.037$, yielding $\dms = 44.07
\pm 0.19$ mag (for comparison, $\dmLCDM = 44.45$ mag). However, as
with PSNID, the SNANA SALT2 fit requires an unacceptably large
$\chi^2/\nu = 57.8/14 = 4.13$. These results do not change
substantially if we use the restricted-wavelength version of SALT2;
the fit remains poor.  The MLCS2k2 results were much better
($\chi^2/\nu = 29.8/14 = 2.13$), with best-fit light-curve parameters
of $A_V = 0.389 \pm 0.193$ mag, $\Delta = -0.011 \pm 0.244$, and $\dmm
= 43.97 \pm 0.14$ mag. Using the conversion equations,
the MLCS2k2 fit $A_V$ corresponds to SALT2 $c = 0.059 \pm 0.091$, and
the SALT2 $x1$ predicts MLCS2k2 $\Delta = -0.201 \pm 0.093$. Interestingly, the MLCS2k2 and SALT2 parameters are similar, but the SALT2 fit is significantly worse.

To summarize, our classification analysis of SN CLA11Tib is
inconclusive. The light curve was best fit by CC~SN templates with STARDUST, while PSNID preferred a SN~Ia fit. Similar to SN CLO12Car, we corrected the light curve for a varying amount of lensing (from 1 mag of demagnification to 3 mag of magnification) and fit with PSNID to see if the luminosity prior was driving the classification. The value of $\PIa$ varied greatly, with the probability dropping to 0 with magnifications $>$ 1.5 mag.  Because of these conflicting results, we further explore both CC~SN and SN~Ia models in our lensing analysis of SN CLA11Tib (\S \ref{sec:lensing-tib}).

\section{Gravitational Lensing Magnification} \label{sec:lensing}

In this section, we derive the magnification of each SN in two
independent ways. First, we use the SNe themselves, from the observed
SN brightness and light-curve fits described in \S
\ref{sec:class}. Second, we independently determine the expected
magnification from cluster mass maps derived by the CLASH team from
strong+weak lensing features.

One approach to inferring the SN magnification is simply measuring the
offset between the inferred distance modulus and that predicted by the
cosmological model ($\dmLCDM$). In that case, however, statistical and
systematic uncertainties in the cosmological parameters (which are
themselves derived in part from SNe~Ia over a wide redshift range)
need to be included in the analysis. A more direct approach is to
compare the CLASH SNe directly to a ``field'' sample at similar
redshifts. For a sufficiently small redshift range, the relation
between distance modulus and redshift can be taken to be linear. The
field sample of SNe~Ia define the zeropoint and slope of that
relationship, and we derive the inferred magnification of our SN~Ia
relative to that locus, with all of these SNe analyzed in an identical
way. This differential approach is insensitive to zeropoint offsets
between the fitters and cosmological parameters, and only makes the
assumption that all SNe~Ia at nearly the same redshift are
consistently standardizable (so it is less susceptible to evolution or
other SN~Ia systematics that vary with redshift).

For the predicted magnification from the CLASH cluster mass maps, we
analyzed both the strong and weak lensing features in the
clusters. The identification of multiple images in the cluster fields
was done using the 16-band \emph{HST} imaging obtained in the context
of the CLASH campaign.  The \citet{Zitrin2009} method revealed nine
multiple image systems in the case of Abell~383 and seven systems in
MACS~J1720. Unfortunately, no multiple-image systems were found in the
field of RX~J1532, the cluster with the lowest total mass in this
study. Where available, spectroscopic redshifts from the literature or
the CLASH-VLT program \citep{Balestra2013} determined the distance to
the strong lensing features. The reliable 16-band photometric
redshifts \citep{Jouvel2013} from the CLASH pipeline delivered the
redshifts of systems without spectroscopic confirmation.  In the case
of weak lensing, we used 5-band CLASH Subaru imaging to derive weak
lensing shear catalogues. The underlying pipeline for image reduction,
shape measurement, and background selection is thoroughly outlined by
\citet{Umetsu2012} and \citet{Medezinski2013}.

To combine the constraints from the weak and strong lensing regime
into a consistent lens model, we used SaWLens \citep{Merten2009,
  Meneghetti2010, Merten2011, Umetsu2012, Medezinski2013}. This
nonparametric method also reconstructs the mass distribution in the
regime far away from the strong lensing features, or where no strong
lensing features are available, and both cases apply in our study.  By
nonparametric, we mean that no {\it a priori} assumptions on the mass
distribution of the lens were made. The method reconstructs the
lensing potential, the rescaled and line-of-sight integrated
counterpart of the Newtonian potential, on an adaptively refined mesh
using both weak and strong lensing inputs.  In the case of RX J1532,
where no strong lensing features were found, we use only the shear
catalogues to constrain the lens mass distribution.  From the lensing
potential maps we directly derive the magnification
\citep[e.g.,][]{Bartelmann2010} at the redshift of interest and read
off its value at the position of each SN, as described individually
below. In order to assign error bars to these values, we run 1500
realizations of each lens model by bootstrapping the weak lensing
input catalogues and randomly sampling through the allowed redshift
error range of each strong lensing feature \citep{Merten2011}.

The independent magnification estimates, from both methods above, are
summarized in Table~\ref{tab:lens-compare}. We consider each object individually below.
Because we could not conclusively classify SN CLA11Tib, we consider
both the SN~Ia and CC~SN models for it (\S \ref{sec:lensing-tib}).

\subsection{SN CLO12Car} \label{sec:lensing-car}

\begin{figure*}
\begin{center}
\includegraphics[width=5.5in]{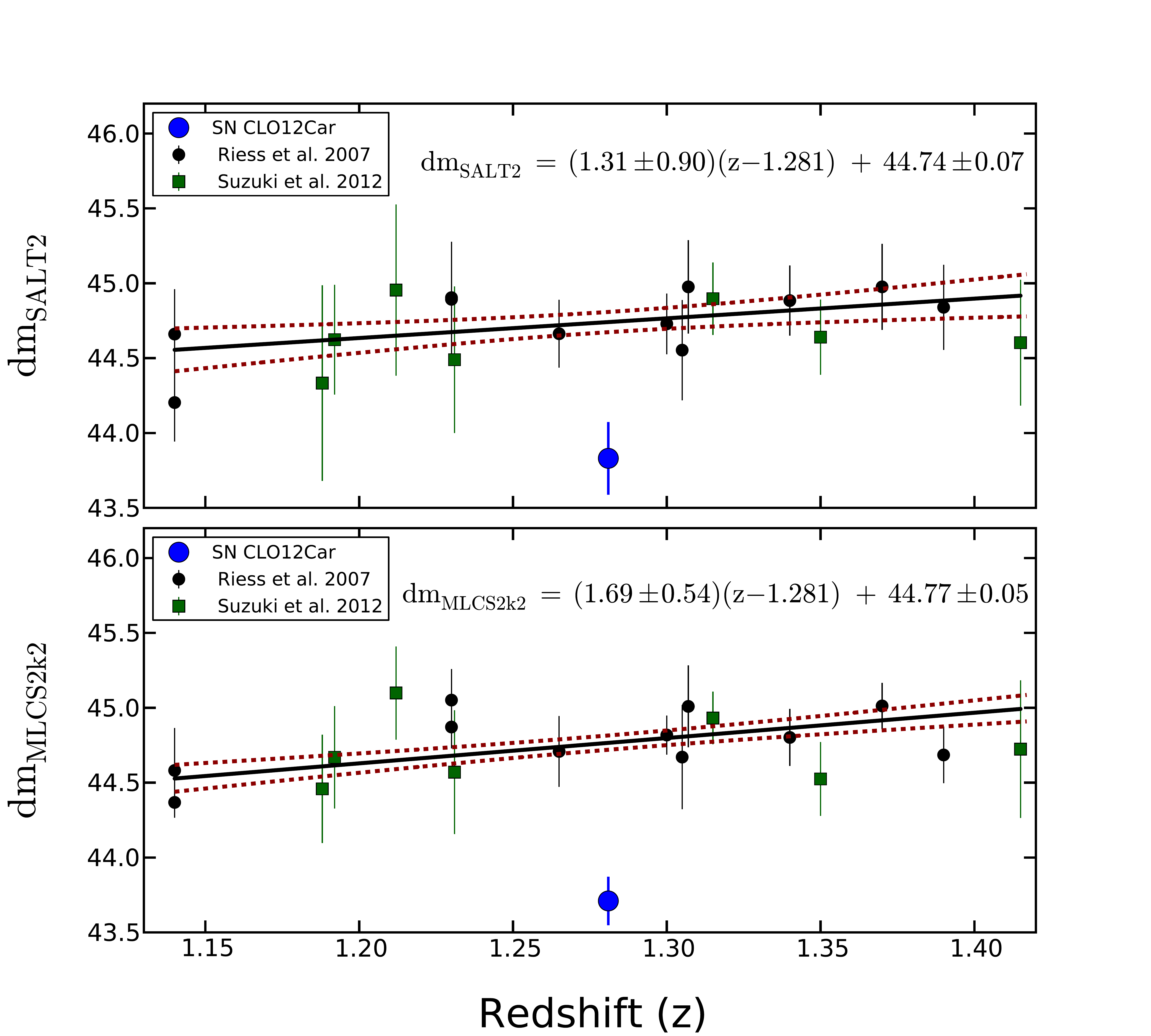}
\caption{Hubble diagrams for SN CLO12Car. The top panel shows SN~Ia
  distances fit with SALT2 and the bottom panel has MLCS2k2 fits. The
  plots contain 18 field SNe~Ia from \citet[][in black]{Riess2007} and
  \citet[][in green]{Suzuki2012} as a comparison sample. Both
  light-curve fitters show that SN~CLO12Car (the large blue point) was
  significantly brighter than typical SNe~Ia at similar redshifts.
  The solid black lines with the displayed equations give the best-fit
  linear model to the comparison sample (empirically approximating a
  cosmological fit over this small redshift range without any
  dependence on cosmological parameters), with the red dashed lines representing the 1$\sigma$ uncertainties. From these data, we predict
  that a typical unlensed SN~Ia at $z = 1.281$ would have $\dms =
  44.74 \pm 0.07$ mag and $\dmm = 44.77 \pm 0.05$ mag. Compared to the
  inferred distance modulus of SN~CLO12Car, we find that it was
  magnified by $\deltams = 0.91 \pm 0.25$ mag or $\deltamm = 1.06 \pm
  0.17$ mag. Both light-curve fitters give a consistent and
  significant magnification (greater than
  unity). \label{fig:car-hubble}}
\end{center}
\end{figure*}

We showed in \S \ref{sec:obs-car} and \S \ref{sec:class-car} that SN
CLO12Car was a SN~Ia at $z=1.281$. In Figure \ref{fig:car-hubble}, we
display two Hubble diagrams for the SN. The top plot contains SN~Ia
distances fit with SALT2 and the bottom plot with MLCS2k2. Both plots
also show a comparison sample of 18 field SNe~Ia from
\citet{Riess2007} and \citet{Suzuki2012} in the range $1.14 \le z \le
1.41$, fit in exactly the same manner as SN~CLO12Car. Consistently
with either light-curve fitter, we find that SN CLO12Car was
significantly brighter than the typical SNe~Ia at a similar
redshift. The solid black line in both panels represents the best
linear fit to the field SN~Ia sample, with the fit slope and intercept
as shown.\footnote{Three of the \citet{Suzuki2012} SNe~Ia plotted in
  Figure \ref{fig:car-hubble} were also behind galaxy clusters and
  potentially lensed. However, the object with the highest estimated
  magnification still only has $\mu = 1.07$, corresponding to $\deltam
  = 0.07$ mag, well within the SN~Ia scatter. Removing these three
  SNe~Ia from the comparison sample has a negligible effect on our
  results.}  From this analysis, we can estimate the lensing
magnification as determined by the SN itself. The linear fits to the
field sample predict $\dms (z = 1.281) = 44.74 \pm 0.07$ mag and $\dmm
(z = 1.281) = 44.77 \pm 0.05$ mag. With these values, we derive that
SN CLO12Car was magnified by $\deltams = 0.91 \pm 0.25$ mag
(corresponding to a magnification $\mu_{\rm SALT2} = 2.31 \pm 0.54$)
in the SALT2 model, or $\deltamm = 1.06 \pm 0.17$ mag ($\mu_{\rm
  MLCS2k2} = 2.65 \pm 0.42$) with MLCS2k2.

\begin{figure*}
\begin{center}
\includegraphics[width=5in]{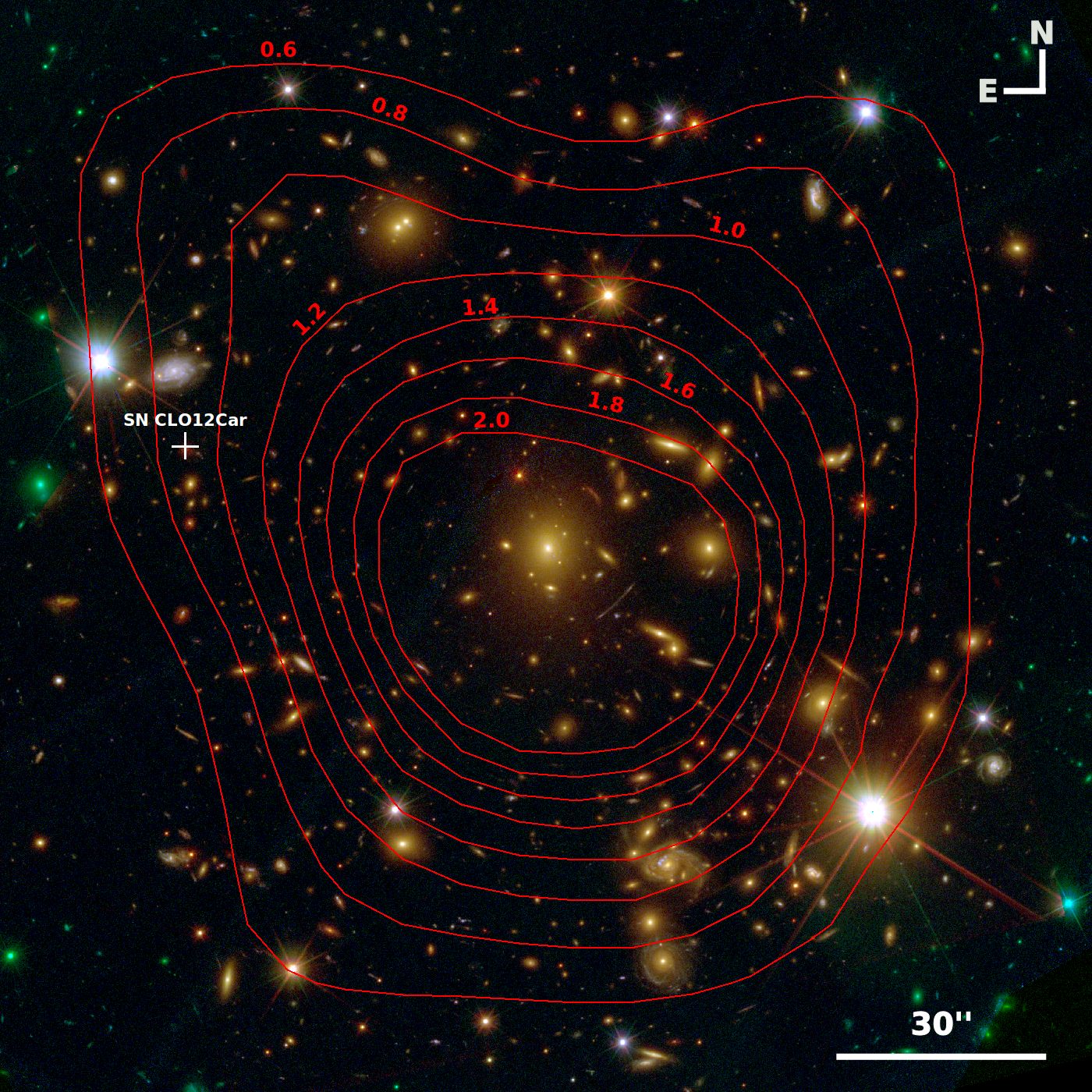}
\caption{Magnification map for SN CLO12Car ($z = 1.281$) behind
  MACSJ1720.2$+$3536 ($z = 0.391$). The image shows an RGB false-color
  background based on 12-band {\it HST}/CLASH optical and near-IR
  images of the cluster field.  The location of the SN is marked with
  a white cross. The contours show the magnitude increase induced by
  the lensing magnification ($\deltam = 2.5\, \log_{10} \mu$) of the
  cluster for a source at the SN redshift. The lensing magnification
  was derived from weak and strong lensing constraints jointly, and
  computed using the SaWLens lensing reconstruction algorithm
  \citep{Merten2009,Merten2011}. Multiple images used for the strong
  lensing constraints in this system will be presented by Zitrin et
  al.~(in prep.). \label{fig:car-map}
}

\end{center}
\end{figure*}

The cluster lensing magnification map (derived as described above)
around MACS~J1720 for a source at the SN redshift ($z = 1.281$) is
shown in Figure \ref{fig:car-map}. From the map, we derive a lensing
magnification for SN CLO12Car of $\mu = 2.15 \pm 0.33$, which
corresponds to $\deltam = 0.83 \pm 0.16$ mag. The magnification from
the light-curve fits and the lensing maps agree to well within
1$\sigma$. Moreover, SN CLO12Car is inconsistent with being unlensed;
its SN~Ia fit requires $\deltam > 0$ at $\sim 4\sigma$.

\subsection{SN CLN12Did} \label{sec:lensing-did}

\begin{figure*}
\begin{center}
\includegraphics[width=5.5in]{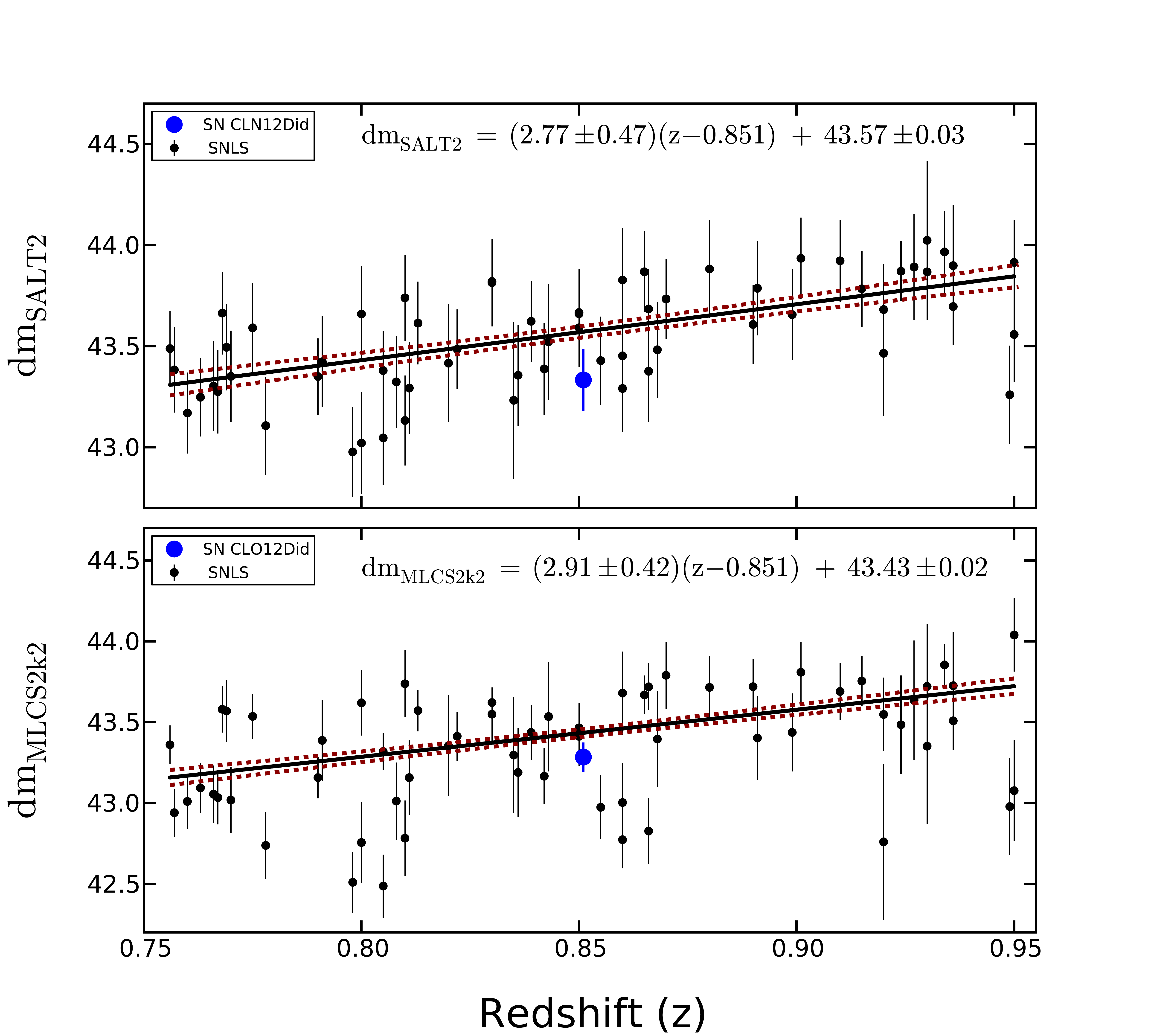}
\caption{Hubble diagrams for SN CLN12Did. As in Figure
  \ref{fig:car-hubble}, the top and bottom panels show SALT2 and
  MLCS2k2 fits, respectively. The field comparison sample consists of
  63 SNe~Ia from SNLS \citep{Guy2010}. Both light-curve fitters
  suggest slight (but insignificant) magnification for the SN, with
  $\deltams = 0.24 \pm 0.15$ mag and $\deltamm = 0.15 \pm 0.09$
  mag. \label{fig:did-hubble}}
\end{center}
\end{figure*}

We showed in \S \ref{sec:obs-did} and \S \ref{sec:class-did} that SN
CLN12Did was a SN~Ia at $z=0.851$. In Figure \ref{fig:did-hubble}, we
display two Hubble diagrams for the SN, mirroring the analysis for
SN~CLO12Car (see \S \ref{sec:lensing-car} and Figure
\ref{fig:car-hubble}). The comparison field sample contains 63 SNe~Ia
from the SNLS sample \citep{Guy2010} in the range $0.75 \le z \le
0.95$. Unlike the case for SN~CLO12Car, we see that SN CLN12Did falls
well within the normal scatter for typical SNe~Ia in the redshift
range. Both light-curve fitters do find SN~CLN12Did to be brighter
than expected compared to the field sample, but not significantly so
($\sim 1\sigma$). Our inferred magnification estimates for SN~CLN12Did
are $\deltams = 0.24 \pm 0.15$ mag ($\mu_{\rm SALT2} = 1.25 \pm 0.18$)
and $\deltamm = 0.15 \pm 0.09$ mag ($\mu_{\rm MLCS2k2} = 1.15 \pm
0.10$).

\begin{figure*}
\begin{center}
\includegraphics[width=4.3in]{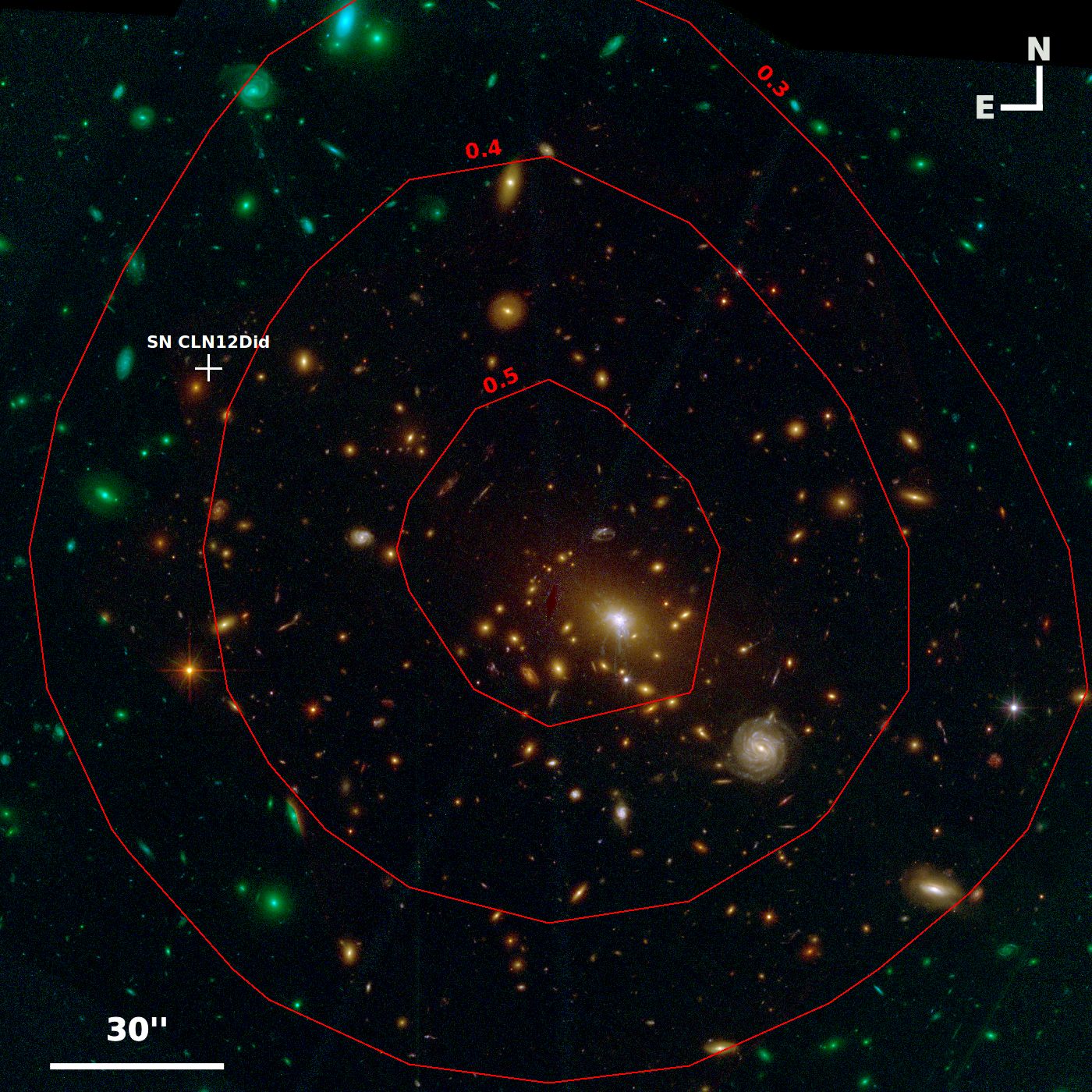}
\caption{Magnification map for SN CLN12Did ($z = 0.851$) behind
  RXJ1532.9$+$3021 ($z = 0.345$). Similar to Figure \ref{fig:car-map},
  the color image is based on CLASH optical and near-IR images, with
  the SN location marked by a white cross and lensing magnification
  ($\deltam$, in mag units) shown with red contours. Because RX~J1532
  is a much weaker lensing cluster than either MACS~J1720 or
  Abell~383, this map was derived using weak lensing constraints
  alone, computed with the SaWLens algorithm \citep{Merten2009,
    Merten2011}. The resolution of the reconstructed map from
  weak-lensing constraints is much poorer than for the other two
  clusters, so the displayed contours are heavily smoothed, and the
  Monte Carlo magnification estimate at the SN position is actually
  $\deltam = 0.28 \pm 0.08$ mag. \label{fig:did-map}}
\end{center}
\end{figure*}

The cluster lensing magnification map around RX~J1532 for a source at
$z = 0.851$ is shown in Figure \ref{fig:did-map}. It predicts a
lensing magnification for SN~CLN12Did of $\deltam = 0.28 \pm 0.08$ mag
($\mu = 1.29 \pm 0.09$).  Just like in the case of SN~CLO12Car, the
lensing map prediction is consistent with the magnification inferred
from the SN light curve, though for SN~CLN12Did the magnification is
only detected at low significance.

\subsection{SN CLA11Tib} \label{sec:lensing-tib}

\begin{figure*}
\begin{center}
\includegraphics[width=4.3in]{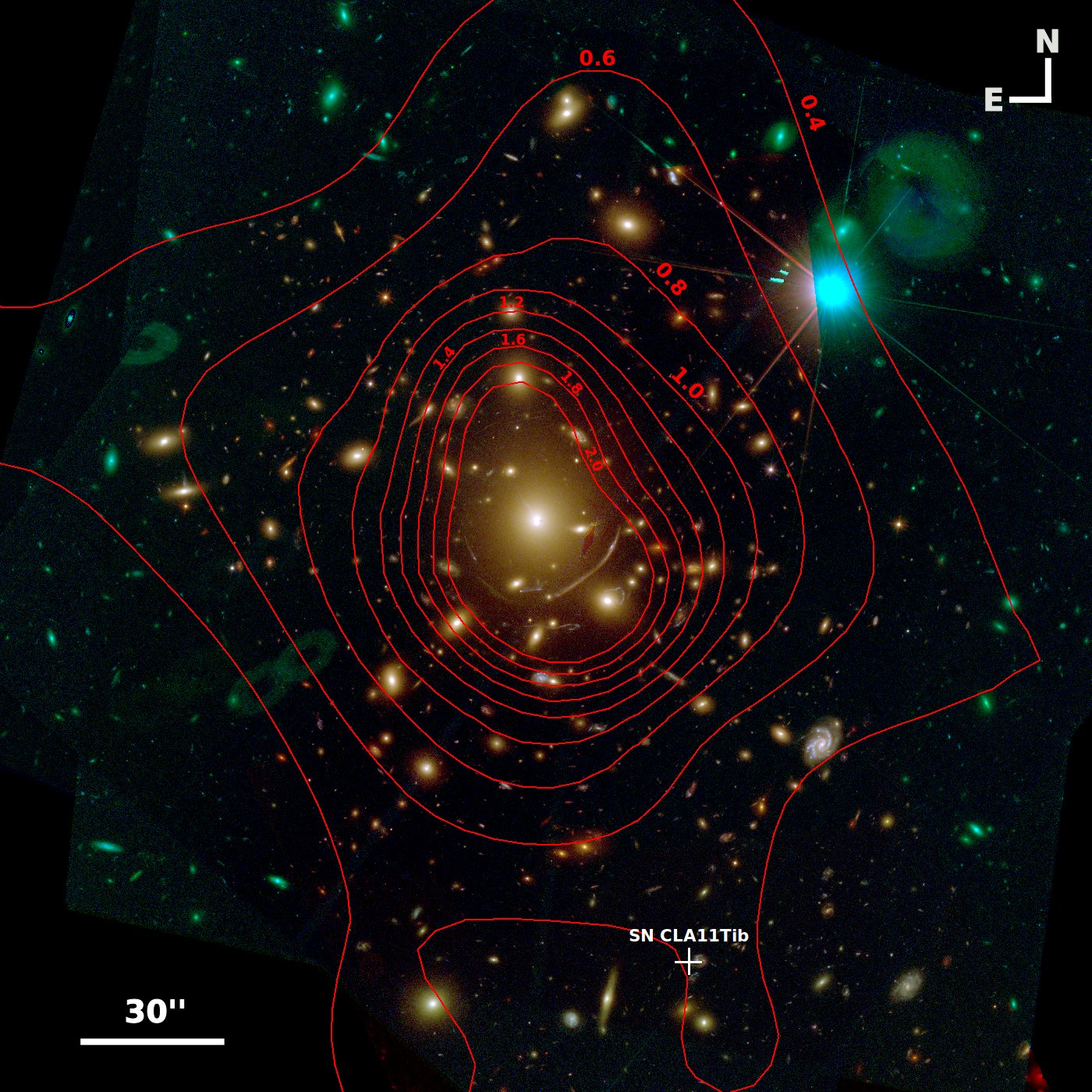}
\caption{Magnification map for SN CLA11Tib ($z = 1.143$) behind Abell
  383, similar to Figures \ref{fig:car-map} and \ref{fig:did-map},
  computed from the SaWLens algorithm \citep{Merten2009, Merten2011}
  from strong and weak lensing constraints. The strong lens multiple
  images in this system were presented by
  \citet{Zitrin2011}. \label{fig:tib-map}}
\end{center}
\end{figure*}

We were unable to classify SN CLA11Tib conclusively (see \S
\ref{sec:class-tib}), so we take a slightly different approach to the
analysis for this object.  Figure \ref{fig:tib-map} shows the
predicted lensing magnification map for a source at the redshift of
SN~CLA11Tib ($z = 1.143$) behind Abell 383. From the map, we derive a
lensing magnification for the SN of $\mu = 1.48 \pm 0.15$, which
corresponds to $\deltam = 0.43 \pm 0.11$ mag.

If we assume that SN~CLA11Tib was a SN~Ia,
repeating the analysis above (as for SN~CLO12Car and SN~CLN12Did)
yields magnification estimates for SN~CLA11Tib of $\deltams = 0.52 \pm
0.20$ mag and $\deltamm = 0.64 \pm 0.15$ mag. However, as mentioned in
\S \ref{sec:class-tib}, the SALT2 light-curve fit is unacceptable,
with $\chi^2/\nu = 57.8/14 = 4.13$, making the inferred distance unreliable. The MLCS2k2 light-curve fit ($\chi^2/\nu = 29.8/14 = 2.13$) is much better, and agrees well with the lensing map prediction (within 1$\sigma$). However, due to the discrepancy of the goodness of fit between MLCS2k2 and SALT2, we hesitate to definitively classify the SN as a SN~Ia.

\begin{figure*}
\begin{center}
\includegraphics[width=6in]{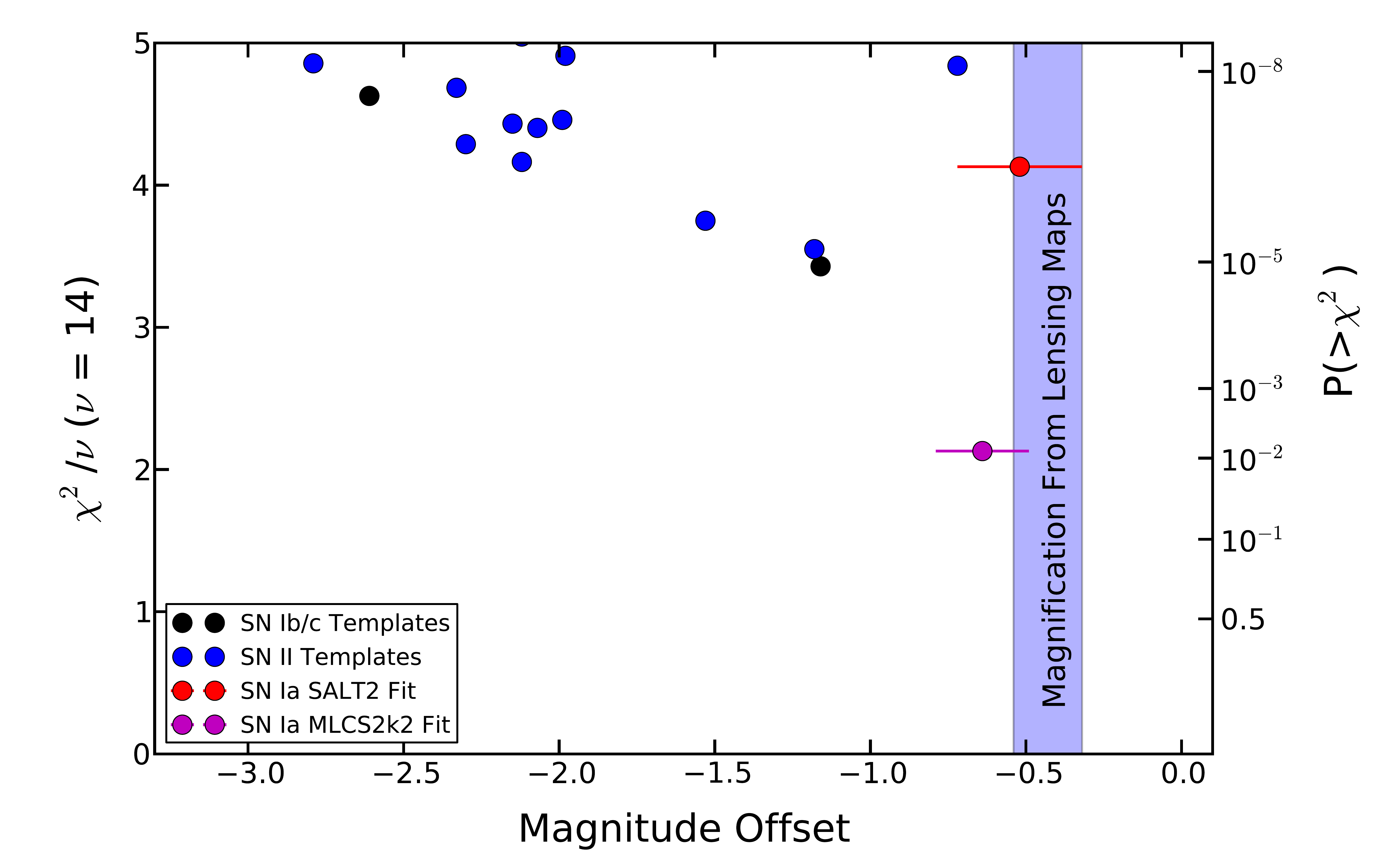}
\caption{Reduced $\chi^2$ (with 14 degrees of freedom) versus
  magnitude offset for SN CLA11Tib. The magnitude offset is a proxy
  for $\deltam$ from the light-curve fit model, as described in the
  text. The cumulative probability for a given $\chi^2_\nu$ value is
  displayed on the right-hand side. The blue shaded region corresponds
  to the 1$\sigma$ range of $\deltam$ derived from the lensing map in
  Figure \ref{fig:tib-map}. The black and blue points correspond to
  SN~Ib/c and SN~II models, respectively, while the red (SALT2) and
  magenta (MLCS2k2) points are the best-fit SN~Ia models. Only the MLCS2k2 fit
  is consistent with both the SN light curve and the
  lensing map magnification prediction, but the CC~SN templates do not
  span the full range of CC~SN luminosities and light
  curves. \label{fig:tib-chisq}}
\end{center}
\end{figure*}

To compare both SN~Ia and CC~SN models in the lensing analysis for
SN~CLA11Tib, we show in Figure \ref{fig:tib-chisq} the goodness-of-fit
for both sets of models. The plot gives the best-fit SN~Ia results as
above, as well as SN~II and SN~Ib/c templates (restricted to those
with $\chi^2/\nu < 5$). For the CC~SN templates, STARDUST calculates
a magnitude offset $\Delta m$ (shown also in Figure \ref{fig:tib-lc})
that is required for the best match of the template to the data, and
we show this on the abscissa of Figure \ref{fig:tib-chisq}.

We found that the best-fitting
CC~SN models (albeit at $\chi^2/\nu > 3$)  from STARDUST were the SN~Ic SDSS-017548 and the SN~IIP
SDSS-018793. The SN~Ic and SN~IIP models had best-fit model $m_J =
23.52 \pm 0.08$ mag and $m_J = 23.48 \pm 0.08$ mag, respectively. The
observed red color of SN~CLA11Tib matches the SN~Ic model with no
relative extinction, but the SN~II model requires $A_V = 0.7$
mag. Including this, we find the unextinguished absolute magnitudes of
the models are $M_R = -19.34 \pm 0.08$ mag (SN~Ic) or $M_R = -20.43
\pm 0.08$ mag (SN~IIP).  These are quite bright for CC~SNe, though
some of this could be a result of the lensing magnification. The peak
$M_R$ required by both models is much brighter than the original
templates; the SN SDSS-017548 and SN SDSS-018793 spectral energy
distributions have $M_R = -18.00$ mag and $M_R = -18.55$ mag,
respectively. The offset between the template and data ($\Delta m$;
Figures \ref{fig:tib-lc} and \ref{fig:tib-chisq}) is $-1.16$ mag for
the SN~Ic model and $-1.18$ mag for the SN~IIP model. These would be
our best estimates for the lensing magnification based on the SN light
curve, but an abundance of caution is warranted, as these estimates
assume that SN CLA11Tib had the same intrinsic luminosity as the
best-fit template, a dubious proposition given the large luminosity
scatter of CC~SNe \citep{Li2011,Drout2011,Kiewe2012}. Because of this, if we assume a CC~SN fit, we cannot claim that there is any consistency or
inconsistency between the lensing prediction for SN~CLA11Tib and its
brightness.

\section{Discussion and Conclusion} \label{sec:disc}

SN CLO12Car clearly represents our most exciting result. We find that
it is likely a SN~Ia at $z = 1.281$, gravitationally magnified by
$\sim 1.0 \pm 0.2$ mag. The magnification derived from two independent
methods (the SN light curve and the cluster lensing maps), were
consistent to within 1$\sigma$. SN~CLO12Car is the first SN~Ia which
is both measurably magnified (i.e., with $\deltam > 0$ at high
significance) and for which the lensing magnification can be precisely
and independently derived. SN~CLN12Did is also a SN~Ia with a
consistent (but not significant) supernova-based and model-based
lensing magnification, whereas the unclear classification of
SN~CLA11Tib does not allow for a test of the lens model prediction.

The lens models used here were constructed jointly with both
strong-lensing constraints (multiple images) and weak-lensing
constraints.  In principle, the strong-lensing multiple images
directly constrain the deflection field, which is the gradient of the
potential, while the weak lensing ellipticity measurements further out
constrain the (reduced) shear, a combination of second-order
 derivatives of the potential. As an alternative to the analysis presented 
 here (where we independently
 tested the estimated magnifications), lensed standard candles like
SNe~Ia can play an important role as an additional \emph{direct},
local constraint on the magnification (which is a function of the
convergence $\kappa$, and the shear $\gamma$, both
constructed from second-order derivatives of the potential) in the
lens model \citep[e.g.,][]{Riehm2011}.

Gathering large samples of lensed SNe~Ia is clearly of interest. Due
to the magnification power of the foreground lenses, lensed SNe could
be expected to be seen to higher redshifts than field SNe (for a given
limiting magnitude), with applications to extending the redshift
coverage of cosmological parameter measurements \citep[if the lensing
  magnification can be estimated precisely and accurately enough,
  e.g.,][]{Zitrin2013} or SN rate measurements to constrain progenitor
models \citep[e.g.,][]{Graur2014,Rodney2013}.  There is
also the exciting prospect of finding multiply-imaged SNe, with
measured time-delays as a new cosmographic constraint
\citep[e.g.,][]{Holz2001, Goobar2002, Oguri2003a, Oguri2003b}.

SN~Ia luminosity distances, combined with lens model magnification
estimates (which depend on angular diameter distances to and between
the lens and source) or time-delay distances from multiple SN images,
could also open up new tests of general relativity
\citep[e.g.,][]{Daniel2010, Jain2010, Schwab2010} or even fundamental
cosmological assumptions, like the distance duality relation
\citep[e.g.,][]{Lampeitl2010, Liang2013}.

Though the CLASH survey has ended, the three lensed SNe found behind
CLASH clusters herald the promise of many more such discoveries in the
near future, including from wide-field ground-based surveys like the
Dark Energy Survey \citep{Frieman2013, Bernstein2012}, Pan-STARRS \citep[][and
  references therein]{Rest2013}, and ultimately LSST \citep{LSST2009};
ground-based surveys that target massive galaxy clusters, such as
CluLeSS (Jha et al., in prep.) or SIRCLS \citep{Graham2013}; and
targeted space-based surveys like the \emph{HST} Frontier Fields (PI
M.~Mountain; with SN follow-up observations in program GO-13396, PI
S.~Rodney). SNe~Ia were the key to the discovery of the accelerating
Universe; perhaps gravitationally lensed SNe~Ia will play a starring
role in further illuminating the dark Universe, probing not just dark
energy, but dark matter as well.

\acknowledgments

We would like to thank Rick Kessler for invaluable assistance with
SNANA. We are grateful to Jakob Nordin for helpful discussions, and
Saul Perlmutter and the Supernova Cosmology Project for communication
and coordination.

This research at Rutgers University was supported through NASA/{\it
  HST} grant GO-12099.14 and National Science Foundation (NSF) CAREER
award AST-0847157 to S.W.J. Support for S.R.~and A.Z.~was provided by NASA 
through Hubble Fellowship grants HST-HF-51312.01 and HST-HF-51334.01, 
respectively, awarded by the Space Telescope Science Institute, which is 
operated by the Association of Universities for Research in Astronomy, Inc., 
for NASA, under contract NAS 5-26555. L.I.~thanks Basal/CATA CONICYT funding 
for support. This research was carried out in part at the Jet Propulsion Laboratory,
California Institute of Technology, under a contract with NASA.
Support for {\it HST} programs GO-12065 and GO-12099 was provided by
NASA through a grant from the Space Telescope Science Institute, which
is operated by the Association of Universities for Research in
Astronomy, Incorporated, under NASA contract NAS5-26555.  A.V.F.~is
also grateful for the support of NSF grant AST-1211916, the TABASGO
Foundation, and the Christopher R.~Redlich Fund. Support for M.N. was provided in 
part by grant PRIN-INAF 2010.

Some of the observations reported in this paper were obtained with the
Southern African Large Telescope (SALT), through Rutgers University
program 2011-3-RU-001 (PI C.~McCully).  This work was supported by a NASA 
Keck PI Data Award, administered by the NASA Exoplanet Science Institute. Data presented herein were obtained at the W. M. Keck Observatory from telescope time allocated to the National Aeronautics and Space Administration through the agency's scientific partnership with the California Institute of Technology and the University of California. The Observatory was made possible by the generous financial support of the W. M. Keck Foundation.
The authors recognize and acknowledge the very significant
cultural role and reverence that the summit of Mauna Kea has always
had within the indigenous Hawaiian community, and we are most
privileged to have the opportunity to explore the Universe from this
mountain.  This work is based in part on data collected at the Subaru
telescope and obtained from the Subaru-Mitaka-Okayama-Kiso Archive,
which is operated by the Astronomy Data Center, National Astronomical
Observatory of Japan.

Additional data were obtained at the Gemini Observatory, which is
operated by the Association of Universities for Research in Astronomy,
Inc., under a cooperative agreement with the NSF on behalf of the
Gemini partnership: the National Science Foundation (United States),
the National Research Council (Canada), CONICYT (Chile), the
Australian Research Council (Australia), Minist\'{e}rio da
Ci\^{e}ncia, Tecnologia e Inova\c{c}\~{a}o (Brazil) and Ministerio de
Ciencia, Tecnolog\'{i}a e Innovaci\'{o}n Productiva (Argentina).  The
data were taken as part of programs GN-2012A-Q-32 and GN-2013A-Q-25.

\clearpage

\bibliographystyle{apj}
\bibliography{lensed}

\begin{deluxetable*}{cccccc}
\tablecaption{SN CLO12Car Photometry \label{tab:car-phot}}
\tablehead{
\colhead{UT Date} & \colhead{MJD} & \colhead{Instrument} & 
\colhead{Filter} & \colhead{Exposure Time (s)} & \colhead{Magnitude}}
\startdata
2012 May 05 & 56052.58 & ACS & F850LP & 991.0 & $>$ 25.546 \\
2012 May 05 & 56052.64 & WFC3/IR & F160W & 1408.8 & $>$ 25.347 \\
2012 May 09 & 56056.02 & WFC3/IR & F105W & 1408.8 & $>$ 26.016 \\
2012 May 09 & 56056.04 & WFC3/IR & F140W & 1005.9 & $>$ 25.808 \\
2012 May 22 & 56069.68 & ACS & F814W & 1007.0 & $>$ 26.666 \\
2012 Jun. 17 & 56095.66 & ACS & F850LP & 1032.0 & 24.597 $\pm$ 0.130 \\
2012 Jun. 17 & 56095.68 & ACS & F814W & 975.0 & 25.148 $\pm$ 0.082 \\
2012 Jun. 17 & 56095.73 & WFC3/IR & F160W & 1408.8 & 23.451 $\pm$ 0.050 \\
2012 Jun. 17 & 56095.75 & WFC3/IR & F110W & 1005.9 & 23.856 $\pm$ 0.024 \\
2012 Jul. 02 & 56110.09 & WFC3/IR &  F105W & 1005.9 & 23.750 $\pm$ 0.042 \\
2012 Jul. 02 & 56110.15 & WFC3/IR & F140W & 1005.9 & 23.271 $\pm$ 0.025 \\
2012 Jul. 16 & 56124.09 & WFC3/IR & F105W & 1005.9 & 23.791 $\pm$ 0.054 \\
2012 Jul. 16 & 56124.15 & WFC3/IR & F140W & 1005.9 & 23.250 $\pm$ 0.025 \\
2012 Jul. 23 & 56131.31 & WFC3/IR & F160W & 455.9 & 23.340 $\pm$ 0.063 \\
2012 Jul. 23 & 56131.38 & WFC3/IR & F105W & 455.9 & 24.138 $\pm$ 0.056
\enddata
\tablecomments{All supernova photometry is reported in Vega
  magnitudes. The nondetections are reported as 3$\sigma$ upper
  limits.}
\end{deluxetable*}

\begin{deluxetable*}{cccccc}
\tablecaption{SN CLN12Did Photometry \label{tab:did-phot}}
\tablehead{
\colhead{UT Date} & \colhead{MJD} & \colhead{Instrument} & 
\colhead{Filter} & \colhead{Exposure Time (s)} & \colhead{Magnitude}}
\startdata
2012 Feb. 03 & 55960.64 & ACS & F625W & 1032.0 & 26.699 $\pm$ 0.171 \\
2012 Feb. 03 & 55960.65 & ACS & F850LP & 1017.0 & 25.066 $\pm$ 0.150 \\
2012 Feb. 18 & 55975.67 & ACS & F775W & 1032.0 & 23.041 $\pm$ 0.016 \\
2012 Feb. 18 & 55975.69 & ACS & F606W & 998.0 & 24.186 $\pm$ 0.017 \\
2012 Mar. 03 & 55989.84 & ACS & F814W & 1032.0 & 23.175 $\pm$ 0.015 \\
2012 Mar. 03 & 55989.86 & ACS & F435W & 1018.0 & $>$ 28.560 \\
2012 Mar. 03 & 55990.24 & ACS & F625W & 1032.0 & 24.182 $\pm$ 0.023 \\
2012 Mar. 03 & 55990.25 & ACS & F850LP & 1017.0 & 22.929 $\pm$ 0.026 \\
2012 Mar. 03 & 55990.31 & ACS & F475W & 1032.0 & 27.057 $\pm$ 0.138 \\
2012 Mar. 03 & 55990.32 & ACS & F775W & 1013.0 & 23.034 $\pm$ 0.016 \\
2012 Mar. 03 & 55990.37 & WFC3/IR & F110W & 1508.8 & 22.915 $\pm$ 0.013 \\
2012 Mar. 03 & 55990.39 & WFC3/IR & F160W & 1005.9 & 23.041 $\pm$ 0.041 \\
2012 Mar. 16 & 56002.61 & ACS & F606W & 1032.0 & 25.491 $\pm$ 0.038 \\
2012 Mar. 16 & 56002.62 & ACS & F814W & 984.0 & 23.410 $\pm$ 0.018 \\
2012 Mar. 16 & 56002.67 & WFC3/IR & F105W & 1305.9 & 23.138 $\pm$ 0.019 \\
2012 Mar. 16 & 56002.69 & WFC3/IR & F140W & 1305.9 & 23.540 $\pm$ 0.034 \\
2012 Mar. 18 & 56004.74 & ACS & F475W & 1032.0 & $>$ 28.924 \\
2012 Mar. 18 & 56004.75 & ACS & F850LP & 1001.0 & 23.359 $\pm$ 0.037 \\
2012 Mar. 29 & 56015.52 & ACS & F435W & 1032.0 & $>$ 28.625 \\
2012 Mar. 29 & 56015.53 & ACS & F814W & 1017.0 & 24.148 $\pm$ 0.029 \\
2012 Apr. 12 & 56029.62 & ACS & F850LP & 1032.0 & 24.448 $\pm$ 0.079 \\
2012 Apr. 12 & 56029.63 & ACS & F814W & 985.0 & 25.072 $\pm$ 0.058 \\
2012 Apr. 12 & 56029.68 & WFC3/IR & F160W & 1508.8 & 23.316 $\pm$ 0.038 \\
2012 Apr. 12 & 56029.70 & WFC3/IR & F110W & 1005.9 & 23.853 $\pm$ 0.038 
\enddata
\tablecomments{All SN photometry is reported in Vega magnitudes. The
  nondetections are listed as 3$\sigma$ upper limits.}
\end{deluxetable*}

\begin{deluxetable*}{cccccc}
\tablecaption{SN CLA11Tib Photometry \label{tab:tib-phot}}
\tablehead{
\colhead{UT Date} & \colhead{MJD} & \colhead{Instrument} & 
\colhead{Filter} & \colhead{Exposure Time (s)} & \colhead{Magnitude}}
\startdata
2010 Nov. 18 & 55518.91 & ACS & F625W & 1032.0 & $>$ 27.206 \\
2010 Nov. 18 & 55518.92 & ACS & F850LP & 1014.0 & $>$ 25.768 \\
2010 Nov. 18 & 55518.99 & ACS & F775W & 1010.0 & $>$  26.792 \\
2010 Dec. 08 & 55538.41 & ACS & F850LP & 1032.0 & $>$ 25.802 \\
2010 Dec. 08 & 55538.43 & ACS & F814W & 1060.0 & $>$ 26.872 \\
2010 Dec. 28 & 55558.45 & ACS & F435W & 1032.0 & $>$ 27.385 \\
2010 Dec. 28 & 55558.46 & ACS & F814W & 1092.0 & 24.699 $\pm$ 0.049 \\
2011 Jan. 04 & 55565.97 & ACS & F625W & 1032.0 & 25.349 $\pm$ 0.074 \\
2011 Jan. 04 & 55565.98 & ACS & F850LP & 1092.0 & 24.251 $\pm$ 0.091 \\
2011 Jan. 18 & 55579.35 & ACS & F606W & 1032.0 & 26.364 $\pm$ 0.104 \\
2011 Jan. 18 & 55579.36 & ACS & F814W & 1059.0 & 24.394 $\pm$ 0.038 \\
2011 Jan. 22 & 55583.41 & ACS & F775W & 1032.0 & 24.762 $\pm$ 0.066 \\
2011 Jan. 22 & 55583.43 & ACS & F606W & 1073.0 & 26.551 $\pm$ 0.127 \\
2011 Jan. 24 & 55585.08 & WFC3/IR & F105W & 806.0 & 23.755 $\pm$ 0.037 \\
2011 Jan. 24 & 55585.09 & WFC3/IR & F125W & 806.0 & 23.402 $\pm$ 0.036 \\
2011 Jan. 24 & 55585.14 & WFC3/IR & F160W & 906.0 & 23.357 $\pm$ 0.059 \\
2011 Feb. 07 & 55599.38 & ACS & F814W & 1032.0 & 25.082 $\pm$ 0.076 \\
2011 Feb. 07 & 55599.40 & ACS & F435W & 1093.0 & $>$ 27.336 \\
2011 Feb. 21 & 55613.16 & ACS & F850LP & 1994.0 & 25.247 $\pm$ 0.188 \\
2011 Mar. 01 & 55621.43 & ACS & F850LP & 1076.0 & $>$ 25.677 
\enddata
\tablecomments{All SN photometry is reported in Vega magnitudes. The
  nondetections are listed as 3$\sigma$ upper limits. Dolphot was used to derive PSF photometry for the WFC3/IR observations (see text).}
\end{deluxetable*}

\begin{deluxetable*}{cccc}
\tablecaption{Comparing SN Magnifications with Lensing
  Predictions \label{tab:lens-compare}} 
\tablehead{ 
\colhead{} & \multicolumn{2}{c}{Lensing from Light Curve Fits} \\ 
\colhead{SN name} & \colhead{$\Delta m_\mu$ (SALT2)} &
\colhead{$\Delta m_\mu$ (MLCS22k)} & \colhead{$\Delta m_\mu$
  (Lensing Maps)} }
\startdata 
SN CLO12Car & 0.91 $\pm$ 0.25 & 1.06 $\pm$ 0.17 & 0.83 $\pm$ 0.16 \\ 
SN CLN12Did & 0.24 $\pm$ 0.15 & 0.15 $\pm$ 0.09 & 0.28 $\pm$ 0.08 \\
SN CLA11Tib & \nodata & \nodata & 0.43 $\pm$ 0.11 
\enddata 
\tablecomments{Comparison of lensing predictions from light curve fits
  and lensing maps for the three SNe. All of the $\deltam$ table
  entries have units of magnitudes. For both SN CLO12Car and SN
  CLN12Did, there is good agreement (within 1$\sigma$) between both
  lensing methods. The classification of SN CLA11Tib was inconclusive,
  making it difficult to determine a magnification from the light
  curve (see \S \ref{sec:lensing-tib}).}
\end{deluxetable*}

\end{document}